\PassOptionsToPackage{table,xcdraw}{xcolor}
\documentclass[10pt,superscriptaddress,nofootinbib,notitlepage,twocolumn,pra]{revtex4-2}
 \usepackage{algorithm,algorithmic}

\usepackage{mathtools}
\usepackage{amsmath}
\usepackage[shortlabels]{enumitem}

\usepackage{graphicx,epic,eepic,epsfig,amsmath,latexsym,amssymb,verbatim,color}
 
\usepackage{amsfonts}       
\usepackage{nicefrac}       

\usepackage{amsmath}
\usepackage{bbm}

\usepackage{float}
\usepackage{tikz}
\usetikzlibrary{chains}
\usetikzlibrary{fit}
\usepackage{pgflibraryarrows}		
\usepackage{pgflibrarysnakes}		

\usepackage{epsfig}
\usetikzlibrary{shapes.symbols,patterns} 
\usepackage{pgfplots}

\usepackage[strict]{changepage}
\usepackage{hyperref}
\hypersetup{colorlinks=true,citecolor=blue,linkcolor=blue,filecolor=blue,urlcolor=blue,breaklinks=true}

\usepackage[marginal]{footmisc}
\usepackage{url}
\usepackage{theorem}

\newtheorem{proposition}{Proposition}

\newtheorem{theorem}[proposition]{Theorem}

\def\squareforqed{\hbox{\rlap{$\sqcap$}$\sqcup$}}
\def\qed{\ifmmode\squareforqed\else{\unskip\nobreak\hfil
\penalty50\hskip1em\null\nobreak\hfil\squareforqed
\parfillskip=0pt\finalhyphendemerits=0\endgraf}\fi}
\def\endenv{\ifmmode\;\else{\unskip\nobreak\hfil
\penalty50\hskip1em\null\nobreak\hfil\;
\parfillskip=0pt\finalhyphendemerits=0\endgraf}\fi}
\newenvironment{proof}{\noindent \textbf{{Proof~} }}{\hfill $\blacksquare$}

\newcounter{remark}

\newcounter{example}

\mathchardef\ordinarycolon\mathcode`\:
\mathcode`\:=\string"8000
\def\vcentcolon{\mathrel{\mathop\ordinarycolon}}
\begingroup \catcode`\:=\active
  \lowercase{\endgroup
  \let :\vcentcolon
  }

\usepackage{cleveref}
\usepackage{graphicx}
\usepackage{xcolor}

\RequirePackage[framemethod=default]{mdframed}
\newmdenv[skipabove=7pt,
skipbelow=7pt,
backgroundcolor=darkblue!15,
innerleftmargin=5pt,
innerrightmargin=5pt,
innertopmargin=5pt,
leftmargin=0cm,
rightmargin=0cm,
innerbottommargin=5pt,
linewidth=1pt]{tBox}

\newmdenv[skipabove=7pt,
skipbelow=7pt,
backgroundcolor=blue2!25,
innerleftmargin=5pt,
innerrightmargin=5pt,
innertopmargin=5pt,
leftmargin=0cm,
rightmargin=0cm,
innerbottommargin=5pt,
linewidth=1pt]{dBox}
\newmdenv[skipabove=7pt,
skipbelow=7pt,
backgroundcolor=darkkblue!15,
innerleftmargin=5pt,
innerrightmargin=5pt,
innertopmargin=5pt,
leftmargin=0cm,
rightmargin=0cm,
innerbottommargin=5pt,
linewidth=1pt]{sBox}
\definecolor{darkblue}{RGB}{0,76,156}
\definecolor{darkkblue}{RGB}{0,0,153}
\definecolor{blue2}{RGB}{102,178,255}
\definecolor{darkred}{RGB}{195,0,0}

\newcommand{\nc}{\newcommand}
\nc{\rnc}{\renewcommand}
\nc{\lbar}[1]{\overline{#1}}
\nc{\bra}[1]{\langle#1|}
\nc{\ket}[1]{|#1\rangle}
\nc{\ketbra}[2]{|#1\rangle\!\langle#2|}
\nc{\braket}[2]{\langle#1|#2\rangle}

\nc{\proj}[1]{| #1\rangle\!\langle #1 |}
\nc{\avg}[1]{\langle#1\rangle}
\nc{\rank}{\operatorname{Rank}}
\nc{\smfrac}[2]{\mbox{$\frac{#1}{#2}$}}
\nc{\tr}{\operatorname{Tr}}
\nc{\ox}{\otimes}
\nc{\dg}{\dagger}
\nc{\dn}{\downarrow}
\nc{\cA}{{\cal A}}
\nc{\cB}{{\cal B}}
\nc{\cC}{{\cal C}}
\nc{\cD}{{\cal D}}
\nc{\cE}{{\cal E}}
\nc{\cF}{{\cal F}}
\nc{\cG}{{\cal G}}
\nc{\cH}{{\cal H}}
\nc{\cI}{{\cal I}}
\nc{\cJ}{{\cal J}}
\nc{\cK}{{\cal K}}
\nc{\cL}{{\cal L}}
\nc{\cM}{{\cal M}}
\nc{\cN}{{\cal N}}
\nc{\cO}{{\cal O}}
\nc{\cP}{{\cal P}}
\nc{\cQ}{{\cal Q}}
\nc{\cR}{{\cal R}}
\nc{\cS}{{\cal S}}
\nc{\cT}{{\cal T}}
\nc{\cU}{{\cal U}}
\nc{\cV}{{\cal V}}
\nc{\cX}{{\cal X}}
\nc{\cY}{{\cal Y}}
\nc{\cZ}{{\cal Z}}
\nc{\cW}{{\cal W}}
\nc{\csupp}{{\operatorname{csupp}}}
\nc{\qsupp}{{\operatorname{qsupp}}}
\nc{\var}{{\operatorname{var}}}
\nc{\rar}{\rightarrow}
\nc{\lrar}{\longrightarrow}
\nc{\polylog}{{\operatorname{polylog}}}
\nc{\wt}{{\operatorname{wt}}}
\nc{\av}[1]{{\left\langle {#1} \right\rangle}}
\nc{\supp}{{\operatorname{supp}}}

\nc{\argmin}{{\operatorname{argmin}}}

\def\i{\mathbf{i}}

\def\x{\xi}

\nc{\RR}{{{\mathbb R}}}
\nc{\CC}{{{\mathbb C}}}
\nc{\FF}{{{\mathbb F}}}
\nc{\NN}{{{\mathbb N}}}
\nc{\ZZ}{{{\mathbb Z}}}
\nc{\PP}{{{\mathbb P}}}
\nc{\QQ}{{{\mathbb Q}}}
\nc{\UU}{{{\mathbb U}}}
\nc{\EE}{{{\mathbb E}}}
\nc{\id}{{\operatorname{id}}}

\nc{\CHSH}{{\operatorname{CHSH}}}

\nc{\be}{\begin{equation}}
\nc{\ee}{{\end{equation}}}
\nc{\bea}{\begin{eqnarray}}
\nc{\eea}{\end{eqnarray}}
\nc{\<}{\langle}
\rnc{\>}{\rangle}
\nc{\rU}{\mbox{U}}

\nc{\ob}[1]{#1}

\nc{\SEP}{{\text{\rm SEP}}}
\nc{\NS}{{\text{\rm NS}}}
\nc{\LOCC}{{\text{\rm LOCC}}}
\nc{\PPT}{{\text{\rm PPT}}}
\nc{\EXT}{{\text{\rm EXT}}}
\nc{\Sym}{{\operatorname{Sym}}}

\nc{\ERLO}{{E_{\text{r,LO}}}}
\nc{\ERLOCC}{{E_{\text{r,LOCC}}}}
\nc{\ERPPT}{{E_{\text{r,PPT}}}}
\nc{\ERLOCCinfty}{{E^{\infty}_{\text{r,LOCC}}}}
\nc{\Aram}{{\operatorname{\sf A}}}

\usepackage{tikz}
\usepackage{hyperref}
\hypersetup{colorlinks=true,citecolor=blue,linkcolor=blue,filecolor=blue,urlcolor=blue,breaklinks=true}

\makeatletter
\def\grd@save@target#1{%
  \def\grd@target{#1}}
\def\grd@save@start#1{%
  \def\grd@start{#1}}
\tikzset{
  grid with coordinates/.style={
    to path={%
      \pgfextra{%
        \edef\grd@@target{(\tikztotarget)}%
        \tikz@scan@one@point\grd@save@target\grd@@target\relax
        \edef\grd@@start{(\tikztostart)}%
        \tikz@scan@one@point\grd@save@start\grd@@start\relax
        \draw[minor help lines,magenta] (\tikztostart) grid (\tikztotarget);
        \draw[major help lines] (\tikztostart) grid (\tikztotarget);
        \grd@start
        \pgfmathsetmacro{\grd@xa}{\the\pgf@x/1cm}
        \pgfmathsetmacro{\grd@ya}{\the\pgf@y/1cm}
        \grd@target
        \pgfmathsetmacro{\grd@xb}{\the\pgf@x/1cm}
        \pgfmathsetmacro{\grd@yb}{\the\pgf@y/1cm}
        \pgfmathsetmacro{\grd@xc}{\grd@xa + \pgfkeysvalueof{/tikz/grid with coordinates/major step}}
        \pgfmathsetmacro{\grd@yc}{\grd@ya + \pgfkeysvalueof{/tikz/grid with coordinates/major step}}
        \foreach \x in {\grd@xa,\grd@xc,...,\grd@xb}
        \node[anchor=north] at (\x,\grd@ya) {\pgfmathprintnumber{\x}};
        \foreach \y in {\grd@ya,\grd@yc,...,\grd@yb}
        \node[anchor=east] at (\grd@xa,\y) {\pgfmathprintnumber{\y}};
      }
    }
  },
  minor help lines/.style={
    help lines,
    step=\pgfkeysvalueof{/tikz/grid with coordinates/minor step}
  },
  major help lines/.style={
    help lines,
    line width=\pgfkeysvalueof{/tikz/grid with coordinates/major line width},
    step=\pgfkeysvalueof{/tikz/grid with coordinates/major step}
  },
  grid with coordinates/.cd,
  minor step/.initial=.2,
  major step/.initial=1,
  major line width/.initial=2pt,
}
\makeatother

\usepackage{thmtools}
\usepackage{thm-restate}
\usepackage{etoolbox}
\makeatletter
\def\problem@s{}
\newcounter{problems@cnt}

\newcommand{\allproblems}{\problem@s}
\makeatother

\usepackage{amsmath,amsfonts,amssymb,array,graphicx,mathtools,multirow,bm,times,tcolorbox,relsize,booktabs}
\usepackage{appendix}
\usepackage[utf8]{inputenc}
\usepackage[T1]{fontenc}
\usepackage{qcircuit}
\usepackage{xspace}
\usepackage{pgfplots}
\usepackage{bbding}
\usepackage{booktabs}  
\usepackage{amssymb}
\usepackage{makecell}
\usepackage{braket} 
\usepackage{algorithm,algorithmic}

\usepackage{graphicx}
\usepackage{xcolor}
\usepackage[table,xcdraw]{xcolor}
\usepackage{array}
\usepackage{geometry}
\pgfplotsset{compat=1.18}

\newtheorem{assumption}{Assumption}[section]

\definecolor{colortwo}{rgb}{0.4,0.77,0.17}
\definecolor{colorthree}{rgb}{0.01,0.51,0.93}

\definecolor{SeaGreen}{rgb}{0.18, 0.55, 0.34}
\definecolor{CornflowerBlue}{rgb}{0.39, 0.58, 0.93}
\definecolor{RoyalPurple}{rgb}{0.47, 0.32, 0.66}
\definecolor{Aquamarine}{rgb}{0.5, 1.0, 0.83}

\newcommand{\update}[1]{\textcolor{black}{#1}}

\allowdisplaybreaks
\begin{document}

\title{Experimentally validated quantum-secure federated learning \\over a multi-user quantum network}

\author{Zhi-Ping Liu}\thanks{These authors contributed equally.}
\author{Xiao-Yu Cao}\thanks{These authors contributed equally.}
\author{Hao-Wen Liu}\thanks{These authors contributed equally.}
\author{Xiao-Ran Sun}
\author{Yu Bao}
\author{Jian-Yu Shen}
\author{Yu-Shuo Lu}
\affiliation{National Laboratory of Solid State Microstructures and School of Physics, Collaborative Innovation Center of Advanced Microstructures, Nanjing University, Nanjing 210093, China.}
\affiliation{School of Physics and Key Laboratory of Quantum State Construction and Manipulation (Ministry of Education), Renmin University of China, Beijing 100872, China.}
\author{Hua-Lei Yin}\email{hlyin@ruc.edu.cn}
\affiliation{School of Physics and Key Laboratory of Quantum State Construction and Manipulation (Ministry of Education), Renmin University of China, Beijing 100872, China.}
\affiliation{National Laboratory of Solid State Microstructures and School of Physics, Collaborative Innovation Center of Advanced Microstructures, Nanjing University, Nanjing 210093, China.}
\author{Zeng-Bing Chen}\email{zbchen@nju.edu.cn}
\affiliation{National Laboratory of Solid State Microstructures and School of Physics, Collaborative Innovation Center of Advanced Microstructures, Nanjing University, Nanjing 210093, China.}

\date{\today}

\begin{abstract}
Federated learning enables decentralized, privacy-preserving training but remains vulnerable to privacy leakage in the quantum era. Quantum federated learning (QFL) offers a promising path towards enhanced security and efficiency. However, a practical and experimentally validated QFL protocol utilizing near-term quantum techniques to address data privacy has been lacking. Here we present QuNetQFL, a QFL protocol implemented on quantum networks, in which local model updates are masked with distributed quantum secret keys, offering information-theoretic security during aggregation. We experimentally validate the protocol on a four-client quantum network and benchmark its performance using the generated keys on quantum and real-world datasets. Adding a single quantum client significantly improves global accuracy for classifying multipartite entangled and non-stabilizer quantum datasets. For language tasks, we apply QuNetQFL to sentiment analysis by federated fine-tuning of a hybrid classical–quantum language model, achieving comparable and robust performance in simulation and on real quantum hardware. Large-scale simulations further demonstrate scalability to $200$ clients for handwritten-digit recognition, with rapid convergence and a $75\%$ reduction in communication cost via model compression. Our work establishes a practical and scalable route to quantum-secure federated learning for the emerging quantum internet.
\end{abstract}

\maketitle

\section{Introduction}
Deep learning has achieved remarkable success in fields~\cite{lecun2015deep} such as disease diagnosis~\cite{cen2021automatic}, autonomous driving~\cite{kiran2021deep}, and tackling critical scientific challenges~\cite{jumper2021highly}. Notably, large language models~\cite{brown2020language} have demonstrated that model performance improves significantly with larger datasets and model size, following empirical scaling laws~\cite{kaplan2020scaling}. Models like GPT-$4$ derive much of their effectiveness from access to massive public datasets. However, high-quality private data, such as medical and user behavior data~\cite{grishin2019data, price2019privacy}, are often isolated among clients, making centralized training infeasible. 

Federated learning (FL) \textcolor{black}{addresses this challenge by enabling decentralized training over private data, allowing clients to improve a shared model without centralizing sensitive information}~\cite{pmlr-v54-mcmahan17a}. FL \textcolor{black}{has been successfully applied in privacy-sensitive domains}, including healthcare~\cite{rieke2020future}, the Internet of Things~\cite{zhao2020local}, and personalized recommendations~\cite{yang2020federated}. \textcolor{black}{However, although FL avoids direct data sharing, model updates themselves can still leak sensitive information. Recent studies have shown that shared gradients or model updates may expose private data through gradient inversion and related attacks}~\cite{mothukuri2021survey}. \textcolor{black}{This limitation motivates the need for secure aggregation mechanisms that protect local updates while still allowing effective collaborative training.}

Meanwhile, \textcolor{black}{quantum machine learning (QML) has emerged as an active research direction} that combines quantum theory and machine learning \textcolor{black}{and introduces new learning models and paradigms} for quantum information-processing platforms~\cite{biamonte2017quantum, cerezo2022challenges, Zhou_2023}. \textcolor{black}{Building on this direction,} quantum federated learning (QFL)~\cite{ren2023towards} \textcolor{black}{extends federated learning to settings where clients may access quantum resources or use quantum and hybrid quantum-classical models during local training. In this sense, QFL naturally connects the decentralized-learning framework of FL with the emerging learning paradigms of QML.}

\textcolor{black}{However, QFL does not by itself resolve the privacy risks associated with model-update sharing.} Recent protocols \cite{chen2021federated, xia2021quantumfed,liu2025quantum, bhatia2023federated, zhao2023non} have been proposed to improve communication efficiency in QFL, using quantum machine learning models like quantum neural networks (QNNs)~\cite{cerezo2021variational}, which are promising for the noisy intermediate-scale quantum era. While communication efficiency is critical, ensuring data privacy remains the primary concern in QFL. \textcolor{black}{Recent efforts have attempted to mitigate these risks through sophisticated quantum algorithms}~\cite{zhang2022federated, li2024privacy, li2024quantumdelegatedfederatedlearning, wang2024quantum, li2024blind,chu2023cryptoqflquantumfederatedlearning} \textcolor{black}{or by incorporating} differential privacy techniques~\cite{li2021quantum}. However, many of these approaches require extensive quantum resources beyond current capabilities or \textcolor{black}{do not provide information-theoretic privacy guarantees for model-update aggregation}. Consequently, \textcolor{black}{there remains a strong need for} a practical QFL protocol that \textcolor{black}{offers quantum-level security for privacy-preserving aggregation} in the coming era of large-scale quantum computing.

\begin{figure*}[htbp] 
    \centering
    \includegraphics[width=0.9\linewidth]{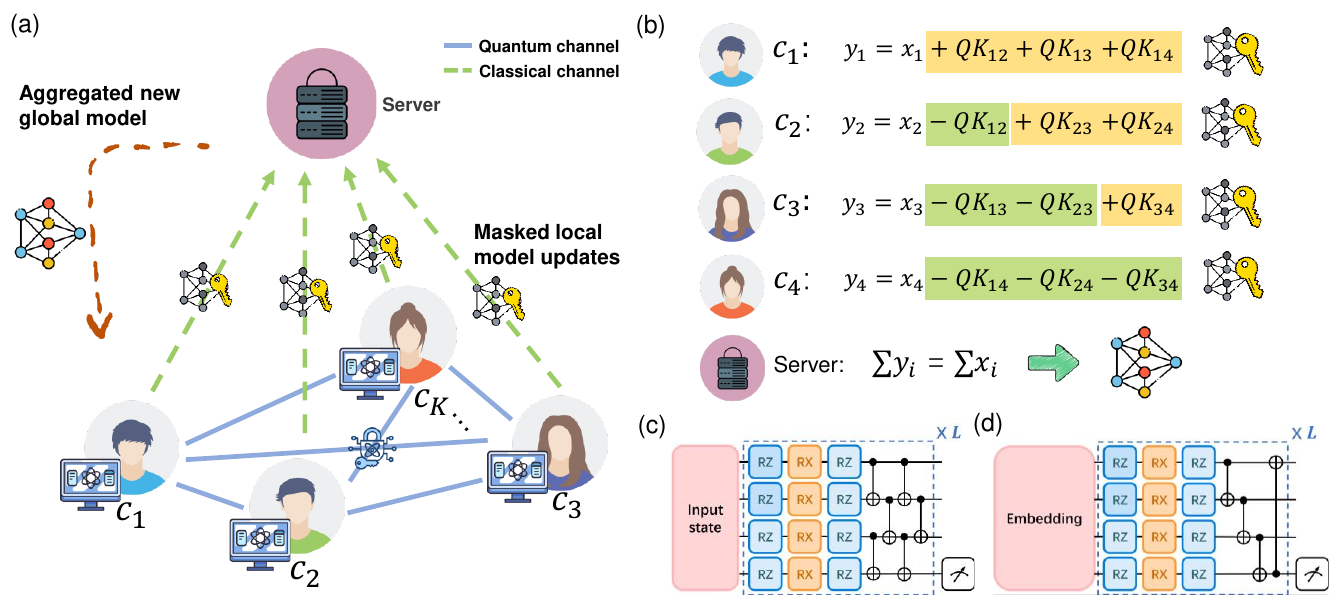}
    \caption{\textbf{Schematic of the QuNetQFL.} (a) The QuNetQFL employs a fully connected quantum network among $K$ clients, where each client uses QKD to securely exchange quantum secure keys with others, enabling pairwise masking of local model updates. Only these masked updates are sent to the server through classical channels, preserving client privacy. The server aggregates the updates to reconstruct the global model parameter, which is then redistributed to all clients. The minimal quantum resource requirement for all clients is access to a fully connected QKD network. (b) Masked secure aggregation in a four-client setup. Here, $\mathbf{x}_i$ represents the raw local updates, and $\mathbf{y}_i$ denotes the masked updates. The server sums the masked updates to obtain the decrypted global model update. (c) and (d) Two types of QNN representations with hardware-efficient ansatzes used in the study.} 
    \label{fig: framework}
\end{figure*}

\textcolor{black}{To address this challenge, a natural direction is to consider security primitives provided by quantum networks.} Quantum key distribution (QKD) \textcolor{black}{, as a foundational primitive of the quantum internet}~\cite{wehner2018quantum}\textcolor{black}{, enables confidential key generation with information-theoretic security}. With \textcolor{black}{major experimental progress in recent years}~\cite{lo2012measurement,lucamarini2018overcoming,xie2022breaking,wang2022twin,gu2022experimental,zhou2023experimental,liu2023experimental,cao2026experimental}, QKD \textcolor{black}{provides a realistic basis for secure aggregation in QFL settings, where both privacy and scalability are important}~\cite{chehimi2023foundations}. \textcolor{black}{More broadly, the rapid development of quantum networks}~\cite{wehner2018quantum,yin2023experimental,azuma2023quantum,xiao2025experimental,xu2025resource} \textcolor{black}{and recent demonstrations of quantum-network capabilities}~\cite{hermans2022qubit,jing2024experimental,zhang2024quantum,schiansky2023demonstration,hua2025experimental,daiss2021quantum,cao2024experimental} \textcolor{black}{further motivate the study of practical privacy-preserving learning protocols built on quantum communication infrastructure.}

In this work, we propose QuNetQFL, a practical quantum federated learning protocol natively implemented on quantum networks, ensuring secure aggregation of model updates with information-theoretic security. Unlike QFL schemes that rely on costly homomorphic encryption~\cite{aono2017privacy, zhang2020batchcrypt, chu2023cryptoqflquantumfederatedlearning} or complex encrypted multi-qubit states preparation and transmission~\cite{zhang2022federated, li2024privacy, li2024quantumdelegatedfederatedlearning, wang2024quantum}, QuNetQFL utilizes secure keys generated by QKD to create pairwise masks for client updates (see Fig.~\ref{fig: framework} for an illustration). Compared to classical secure aggregation schemes~\cite{bonawitz2016practicalsecureaggregationfederated}, our approach realizes information-theoretic one-time-pad masking using QKD-derived keys and, to the best of our knowledge, provides the first experimentally validated quantum-secure federated learning protocol deployed on a multi-user quantum network. Specifically, we validate our protocol by employing four-phase measurement-device-independent (MDI) QKD~\cite{gu2022experimental} in a four-client quantum network. Using a Sagnac interferometer for phase stability, we present a proof-of-principle demonstration of a multi-user quantum network over $6$ kilometers of optical fiber, achieving a secret key rate exceeding $32.8$ kbps. This experimental demonstration confirms that near-term quantum networks can already support practical, scalable distributed learning tasks.


\begin{table*}[ht]
\centering
\caption{\textbf{Summary of key rates (bits per second, bps).} A, B, C, and D represent Alice, Bob, Charlie, and David, respectively.} 
\label{table_res1}
   			\setlength{\tabcolsep}{0.2cm} 
      \renewcommand\arraystretch{1.5}
	
	\begin{tabular}{ cc| c ccc}\hline \hline
    \multicolumn{2}{c|}{3-client}&\multicolumn{4}{c}{4-client}\\ \hline
    Client pair& Key rate &Client pair& Key rate &Client pair& Key rate \\ \hline
            AB &230 kbps &  AB &240 kbps &BC& 44.6 kbps \\ 
            AC &37.6 kbps &  AC &35.6 kbps &BD&36.6 kbps \\ 
            AD &45.9 kbps &  AD &43 kbps &CD& 32.8 kbps \\   
  \hline\hline
	\end{tabular}
\end{table*}


Leveraging experimentally generated quantum keys, we comprehensively benchmark QuNetQFL across a range of scenarios, each utilizing different quantum resources, from quantum, hybrid quantum-classical, to classical models, and quantum to classical datasets. These benchmarks include: (a) classification of two quantum datasets using QNNs; (b) sentiment analysis using a hybrid classical-quantum large language model (Bert-QNN); and (c) handwritten digit recognition using QNNs. In all cases, QuNetQFL demonstrates remarkable performance. We show that adding a quantum client significantly enhances the global model's ability to classify multipartite entangled and quantum magic datasets, resulting in at
least a $2\%$ improvement in test accuracy and a reduction in loss, which highlights the protocol's scalability and robustness. In sentiment analysis, our scheme supports collaborative fine-tuning of Bert-QNN across multiple clients, using widely recognized IMDb, Yelp, and Amazon review datasets~\cite{li2024quantum}. This is validated on both simulated and real quantum hardware, demonstrating the effectiveness of QuNetQFL in enhancing LLM training. For handwritten digit recognition on the MNIST dataset, we achieve comparable accuracy (within $1\%$) under both non-independent and identically distributed (non-IID) and IID data distributions. Further simulations highlight QuNetQFL's scalability to $200$ clients, using classical learning models, LeNet-5, with rapid convergence and an at most $75\%$ reduction in communication cost facilitated by advanced model compression techniques. Overall, our work addresses the critical need for a practical QFL framework that efficiently operates within quantum hardware limitations, ensuring quantum security from the present day through to the era of large-scale quantum computing.


\begin{figure*}[htbp]
    \centering
    \includegraphics[width=1.04\linewidth]{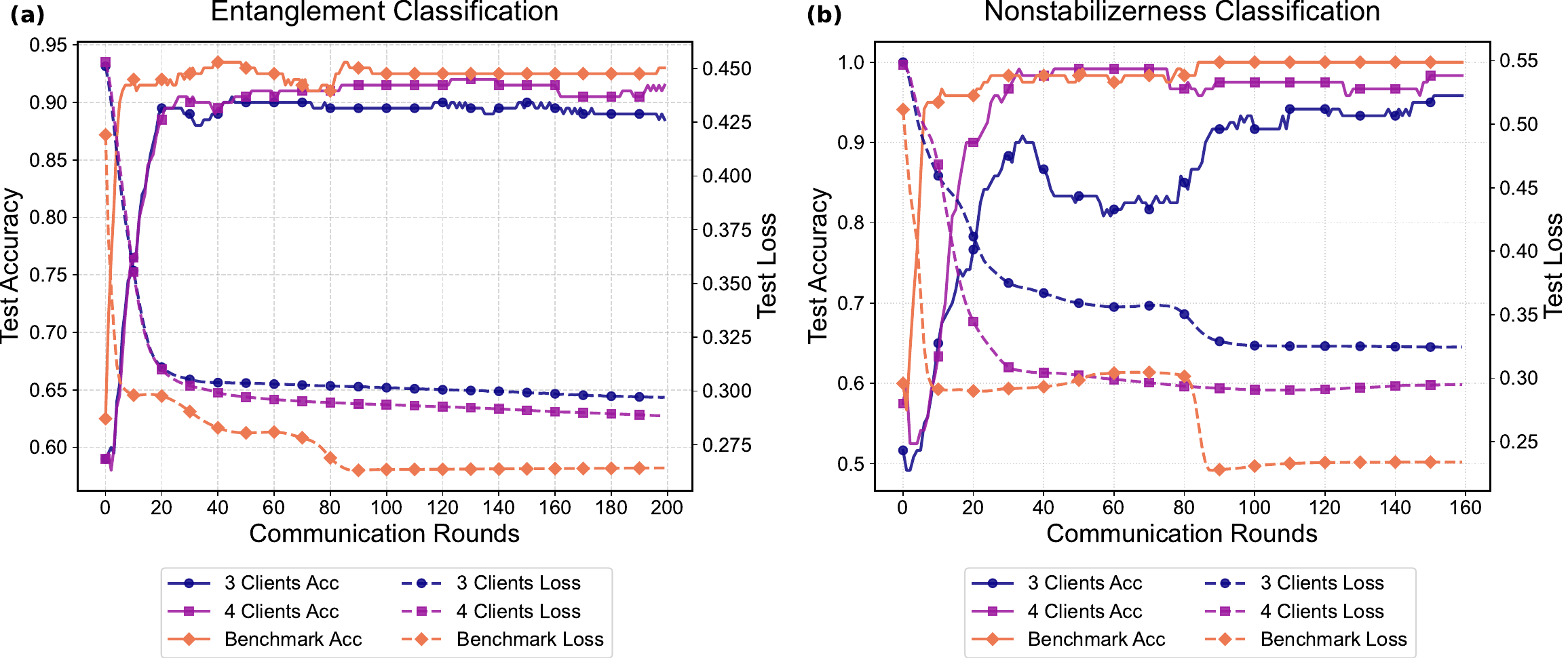}
    \caption{\textbf{Performance of QuNetQFL on quantum state classification using QNN.} 
    (a) Entanglement classification and (b) Nonstabilizerness classification results across different client setups. Adding a single client notably brings the model accuracy and training convergence closer to the benchmark.} 
    \label{fig: CE_Magic} 
\end{figure*} 


\section{Results}\label{Results}
\subsection{Framework of QuNetQFL}
We consider a fundamental QFL setting with one central server and $K$ quantum-enabled clients. Client $k$ holds a local dataset $D_k=\{(\rho_k^i,\mathbf{y}_k^i)\}_{i=1}^{n_k}$, where $\rho_k^i$ is a quantum state and $\mathbf{y}_k^i$ is the corresponding label. Datasets can be native quantum data, or classical data encoded into quantum states $\rho(\mathbf{x}_k^i)$~\cite{larose2020robust}. In communication round $t$, the server broadcasts global QNN parameters $\boldsymbol{\theta}^{t-1}\in\mathbb{R}^M$. Each client then trains a local model $\boldsymbol{\theta}_k^{t}$ by minimizing a local loss $\mathcal{L}(\boldsymbol{\theta}_k^{t})$ and returns the update $\Delta\boldsymbol{\theta}_k^{t}=\boldsymbol{\theta}_k^{t}-\boldsymbol{\theta}^{t-1}$. The server aggregates updates via a weighted sum
\begin{equation}
\Delta\boldsymbol{\theta}^{t}=\sum_{k=1}^K \frac{n_k}{N}\,\Delta\boldsymbol{\theta}_k^{t},\quad
N=\sum_{k=1}^K n_k,
\end{equation}
and updates the global model $\boldsymbol{\theta}^{t}=\boldsymbol{\theta}^{t-1}+\Delta\boldsymbol{\theta}^{t}$. While no raw data leaves clients, sharing gradients or updates is known to be vulnerable to inversion attacks. Ensuring privacy therefore requires encryption methods that remain secure even in the presence of quantum computers. We thus adopt a quantum-enhanced secure aggregation protocol utilizing QKD-based masking, namely QuNetQFL, to achieve information-theoretic security.

\subsection{Quantum-secure aggregation via QKD masking}

QuNetQFL uses pairwise quantum secret keys distributed over a fully connected quantum network to one-time-pad mask local updates (Fig.~\ref{fig: framework}b). In round $t$, the server selects clients $\mathcal{S}^{t}\subseteq\{1,\dots,K\}$ randomly. Each selected pair $(i,j)$ shares a $q$-bit key vector $\mathrm{QK}_{i,j}^{t}\in\mathbb{Z}^{M}$ (symmetric: $\mathrm{QK}_{i,j}^{t}=\mathrm{QK}_{j,i}^{t}$). To make efficient use of these keys, model updates are quantized to $q$ bits before masking (quantizer and de-quantizer are defined in Methods). By choosing $q$ appropriately, this quantization not only enables effective masking but also reduces both key consumption and communication overhead.

Each client uploads a masked, quantized update
\begin{equation}
\Delta\widetilde{\boldsymbol{\theta}}_{i}^{t}
=\Big[\,Q^{q}\!\big(p_i^{t}\cdot\Delta\boldsymbol{\theta}_{i}^{t}\big)+\mathbf{m}_i^{t}\,\Big]\bmod 2^{q},
\end{equation}
where $Q^q(\cdot)$ denotes $q$-bit quantization, $p_i^t$ is the weight proportional to the client’s data size, and $\mathbf{m}_i^t$ is the masking term constructed as a signed sum of pairwise QKD keys shared between client $i$ and the other participants. When all updates are summed, the masks cancel out (see Methods for details), leaving only the aggregated, quantized update.
\begin{equation}
\Delta\boldsymbol{\theta}^{t}
=\Big[\sum_{i\in\mathcal{S}^{t}}Q^{q}\!\big(p_i^{t}\cdot\Delta\boldsymbol{\theta}_{i}^{t}\big)\Big]\bmod 2^{q}.
\end{equation}
The server then reconstructs the global model by de-quantization,
\begin{equation}
\boldsymbol{\theta}^{t}=\boldsymbol{\theta}^{t-1}+D^{q}\!\left(\Delta\boldsymbol{\theta}^{t}\right),
\end{equation}
where $D^q(\cdot)$ reverses quantization. This protocol ensures that only the aggregated update is revealed, while individual contributions remain hidden. Key cost grows as $\mathcal{O}(K^{2}M)$ with client number $K$ and model size $M$, but is substantially reduced by $q$-bit quantization and, at larger scale, by model compression (see Table~\ref{table: quantization}). Full protocol details and pseudocode are provided in Methods (Algorithm~\ref{algo:qunetqfl}).

This quantum-secured aggregation approach provides
perfect privacy against gradient attacks, supporting secure and scalable QFL across distributed networks. The theoretical convergence analysis for QuNetQFL is provided in the Supplementary Materials, offering \textcolor{black}{convergence guarantees and practical guidance for parameter selection and training}.



\begin{figure}
    \centering\includegraphics[width=1.\linewidth]{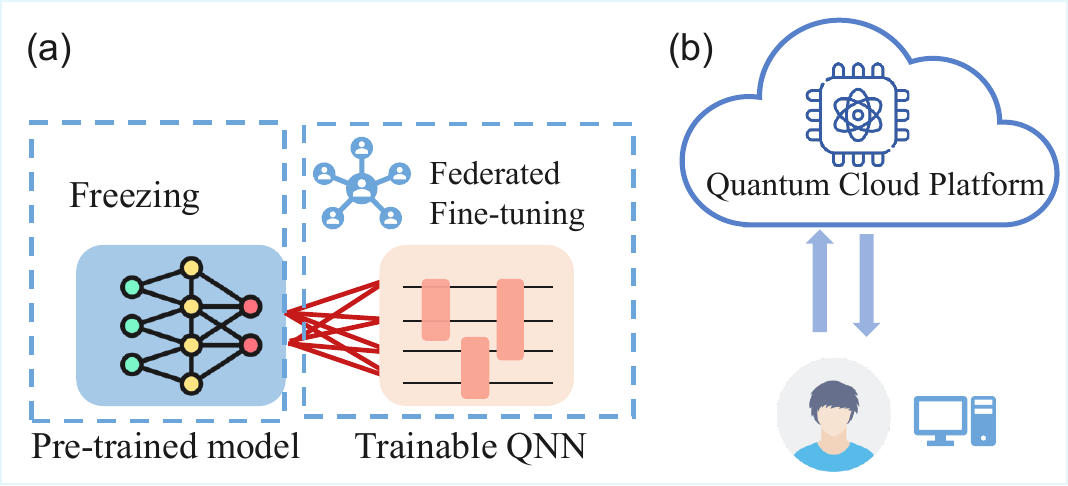}
    \caption{\textbf{Schematic of federated fine-tuning the hybrid BERT–QNN model}. (a) Multiple clients collaboratively fine-tune a hybrid BERT–QNN model on specific datasets, such as sentiment analysis tasks. The pre-trained BERT model is frozen, and only the parameters of the QNN (orange) and the connecting linear layer (red) are efficiently fine-tuned through the QuNetQFL protocol. (b) The trained hybrid model is evaluated on a superconducting quantum computing cloud platform, with performance metrics derived from real quantum hardware evaluations.}
    \label{fig:fl_hybrid}
\end{figure}

\subsection{Experimental quantum secret keys generation}
To ensure an adequate supply of quantum secret keys, we implemented the four-phase MDI-QKD~\cite{gu2022experimental} across three and four clients within a quantum network, with each client pair generating keys over a 200-second duration. These quantum keys were then used to secure several collaborative learning tasks demonstrated in the following section, with local model updates encrypted before aggregation. A Sagnac loop was employed to stabilize the phase, and a $6$-km optical fiber was inserted into the loop.

\update{
In the three-client scenario, the accumulated keys were sufficient to support training a model with up to $1434$ parameters, represented as $32$-bit floating-point values, over $200$ communication rounds. In the four-client case, the number of supported parameters decreased slightly to $1393$, due to the additional pairwise connections increasing the demand on the key pool. The secure key rates achieved for different client pairs are summarized in Table~\ref{table_res1}. Further details of the MDI-QKD implementation, including experimental setup and simulation, are provided in Methods.}

\update{
Although we employ MDI-QKD here, the protocol can be substituted by other QKD schemes depending on practical requirements, serving as a proof-of-principle implementation of QFL. Since secret keys can be pre-generated and stored, the time required for key generation is not a limiting factor in performance. Consequently, our framework is highly flexible and exhibits strong compatibility with emerging quantum communication infrastructures, including satellite-based QKD links, trusted-node architectures, and entanglement-based networks, paving the way toward scalable and secure distributed learning in the quantum era.
}

\begin{figure*}
    \centering 
    \includegraphics[width=0.98\linewidth]{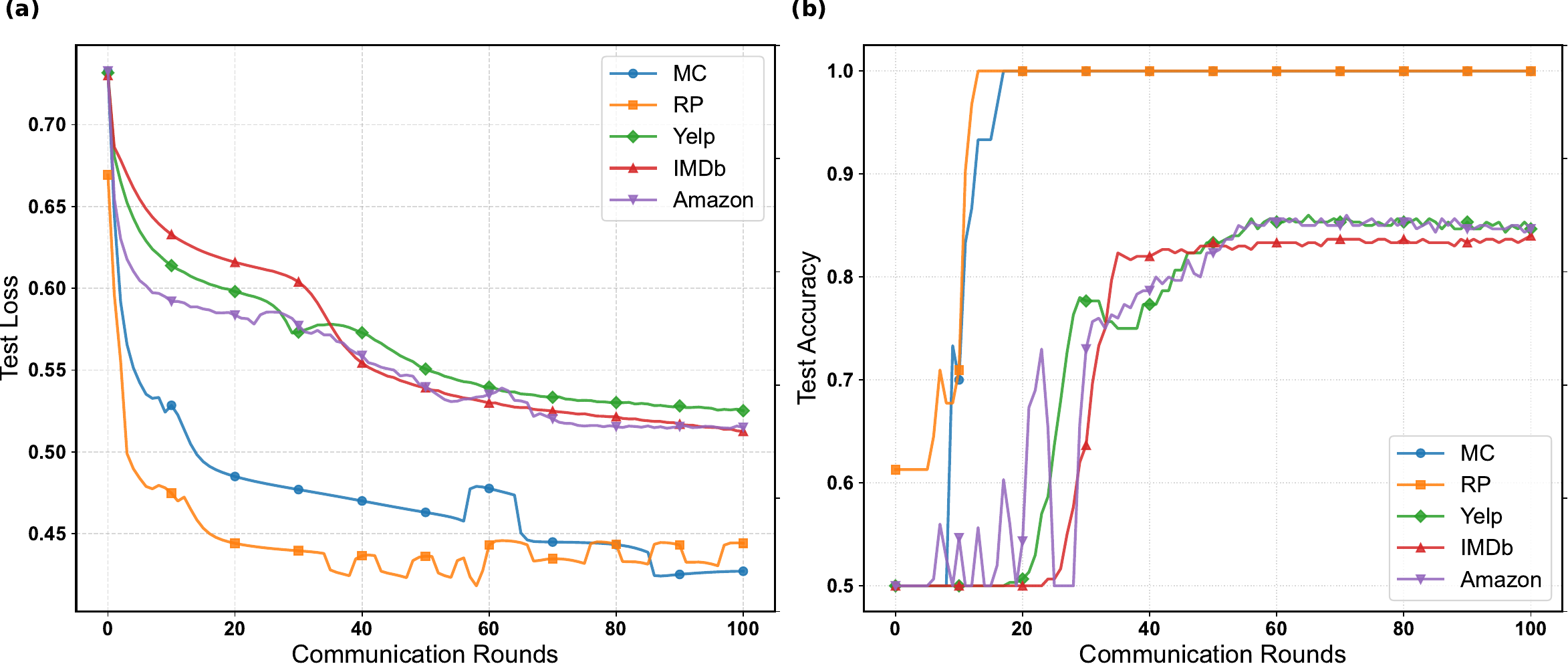}
    \caption{\update{\textbf{Sentiment analysis through Hybrid BERT–QNN}. (a) Training loss curves over $100$ epochs for five language classification tasks on MC, RP, Yelp, IMDb, and Amazon, using a $4$-qubit, $3$-layer HEA QNN. All models are fine-tuned via QuNetQFL with secure aggregation and quantized updates. The loss rapidly decreases within the first $50$ rounds. (b) Corresponding test accuracy curves showing fast convergence. RP and MC converges near‐perfect accuracy by $20$ rounds, while the sentiment tasks stabilize above $84\%$ accuracy.}}
    \label{fig:nlp}
\end{figure*}

\subsection{Classification of quantum datasets using QNNs}

Detecting and certifying quantum resources, such as entanglement and quantum magic (nonstabilizerness), are essential for advancing practical quantum technologies~\cite{Friis_2018, cao2023detection, Zhu_2024, zhu2024ent}. To tackle this, we employed QuNetQFL to classify two quantum datasets: \update{one with multipartite entangled states (high vs. low entanglement) and the other with quantum magic states (non-stabilizer vs. stabilizer)}. We performed these classifications under an IID data distribution setting, utilizing QNNs based on a hardware-efficient ansatz (HEA)~\cite{kandala2017hardware}. Comprehensive details of the datasets and simulations are provided in the Methods. 

By utilizing experimentally generated secret keys for quantum-secure aggregation, we demonstrated secure collaborative learning capabilities across three-client and four-client scenarios. Figure.~\ref{fig: CE_Magic} compares the performance of these scenarios, using $16$-bit quantization ($q = 16$), against an ideal benchmark where all client data is centrally processed. \update{Results show that the final test accuracies for classifying high vs. low entangled states are $88.5\%$ with three clients, $91.5\%$ with four clients, and $93\%$ for the benchmark. For non-stabilizer vs. stabilizer states classification, the final test accuracies are $95.8\%$ with three clients, $98.3\%$ with four clients, and $100\%$ for the benchmark.} In both datasets, the results show a substantial improvement in the global model's performance, \update{\textcolor{black}{with at least a $2\%$ improvement in test accuracy of the four-client setting over the three-client setting, together with a reduction in loss.}} The addition of a single client in both classification tasks brings the model closer to the benchmark, emphasizing the protocol’s scalability and potential for secure, collaborative learning across multiple quantum clients, each operating its own QNN and serving as an individual quantum data center~\cite{Liu_2024}. 


\begin{table*}[t]
\centering
\caption{\textcolor{black}{\textbf{Final test accuracy of BERT-QNN and the dimension-matched classical baseline}}}
\label{table:sim_vs_hw_acc}
\setlength{\tabcolsep}{0.25cm}
\renewcommand\arraystretch{1.3}
{\update{
\begin{tabular}{lccc}
\hline\hline
Task     & BERT-QNN Sim. & BERT-QNN Hardware & \textcolor{black}{Classical Baseline} \\ 
\hline
MC       & 1.000           & 1.000        & \textcolor{black}{1.000} \\ 
RP       & 1.000           & 1.000        & \textcolor{black}{1.000} \\  
Yelp     & 0.847           & 0.850        & \textcolor{black}{0.840} \\  
IMDb     & 0.840           & 0.833        & \textcolor{black}{0.820} \\  
Amazon   & 0.847           & 0.810        & \textcolor{black}{0.817} \\  
\hline\hline
\end{tabular}
}}
\end{table*}

\subsection{Sentiment analysis using Hybrid BERT–QNN}

To further evaluate the scalability and hardware‐relevance of QuNetQFL beyond simulations, we apply our protocol to a series of two-class language classification tasks: MC, RP, two simple synthetic datasets introduced from Ref.~\cite{lorenz2023qnlp} and three sentiment analysis benchmarks from real world (Yelp, IMDb, Amazon). These tasks are tackled by federated fine-tuning the quantum subcircuit of a hybrid BERT–QNN model (see Fig.~\ref{fig:fl_hybrid} for an illustration). In this simulation, four clients participate, each receiving an equal share of the training examples in an IID data partition. The quantum subcircuit consists of a $4$-qubit three‐layer QNN with a HEA. Further simulation details are provided in the Methods and Supplementary Materials. 

Figure.~\ref{fig:nlp} shows the simulation results of test accuracy and loss curves for these five language classification tasks. These results underscore QuNetQFL’s adaptability to hybrid classical–quantum architectures. To further validate end‐to‐end operation on real quantum hardware, we transpile the trained quantum circuits for these tasks into the device’s native gate set and run them on the BAQIS Quafu quantum computing cloud ($156$-qubit superconducting quantum chip with a median 1Q error of $7.6 \times 10^{-4}$ and a median 2Q error of $1.8 \times 10^{-2}$), averaging over $4000$ shots per circuit evaluation. Table~\ref{table:sim_vs_hw_acc} compares the final test accuracy between numerical simulations and real quantum hardware. For synthetic datasets MC and RP, both simulated and quantum hardware results achieved $100\%$ accuracy. The quantum hardware slightly outperformed the simulated result on the Yelp dataset, while an accuracy degradation of $0.7\%$ and $3.7\%$ was observed on IMDb and Amazon, respectively. These results demonstrates the robustness of the trained model to noise on real quantum devices using our method.

\textcolor{black}{To provide a controlled classical reference, we further evaluated a dimension-matched classical baseline under the same frozen BERT encoder, four-client IID partition, $100$-round QuNetQFL protocol, $16$-bit quantization, and QKD-key-based secure aggregation. In this baseline, the $768$-dimensional BERT representation is reduced to a $4$-dimensional feature vector and then processed by a lightweight classical nonlinear classification head. As shown in Table~\ref{table:sim_vs_hw_acc}, the BERT-QNN model achieves comparable final-round performance to this classical baseline, with identical accuracy on MC and RP and slightly higher simulation accuracy on Yelp, IMDb, and Amazon. This comparison is intended as a controlled classical reference rather than evidence of quantum advantage.}

\begin{figure*}[htbp]
    \centering
    \includegraphics[width=1.0\linewidth]{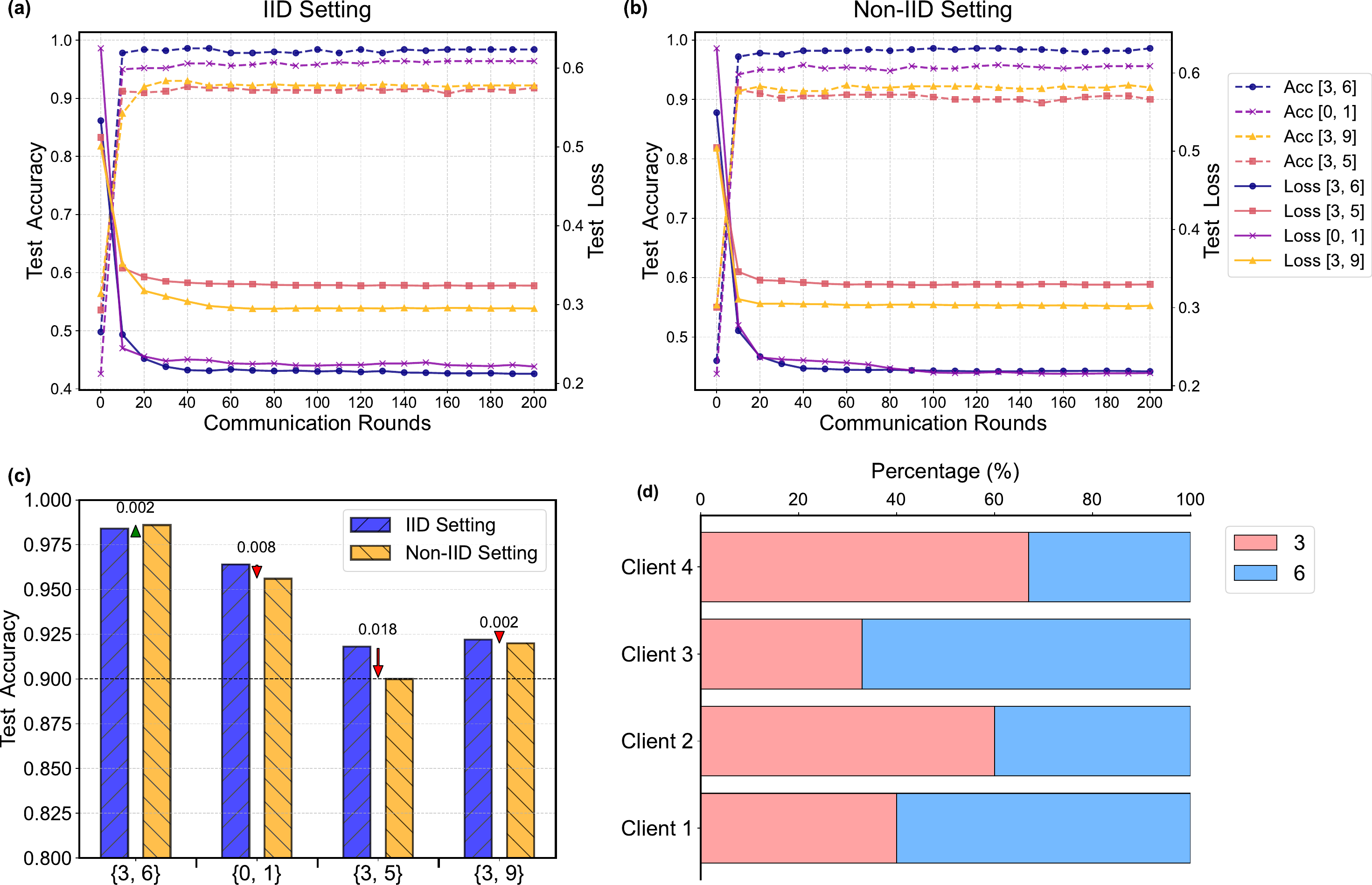}
    \caption{\textbf{Evaluation of QuNetQFL on the MNIST classification task using QNN}. (a) IID setting showing test accuracy and loss across 200 communication rounds for client combinations $\{3,6\}$, $\{0,1\}$, $\{3,5\}$, and $\{3,9\}$ with $16$-bit quantization. (b) Non-IID setting with identical client configurations as (a) for comparative analysis. (c) Final test accuracy comparison between IID and non-IID settings, with annotated accuracy differences illustrating QuNetQFL's robustness across data distributions. (d) Non-IID client data distribution proportions for classes $\{3,6\}$ (same for other cases), highlighting dataset heterogeneity.}
    \label{fig:mnist_iid_noniid}
\end{figure*}

\subsection{Handwriting recognition by QNNs}

To demonstrate the robustness and adaptability of QuNetQFL in real-world, heterogeneous data environments, we evaluated its performance across both IID and non-IID settings for four clients using QNNs on the classical MNIST dataset. We created multiple two-class subsets from MNIST, specifically $\{3,6\}$, $\{0,1\}$, $\{3,9\}$, and $\{3,5\}$, where, for instance, $\{3,9\}$ represents images of digits “$3$” and “$9$”. Each client was allocated $500$ training samples from the MNIST training set, while the server test set comprised $500$ samples from the MNIST test set. In the IID setting, each client’s data contained an equal number of samples from each category. In contrast, the non-IID setting introduced imbalanced distributions to better reflect real-world scenarios where data varies across devices, \update{presenting greater challenges for FL effectiveness.} The data splits among the four clients for two classes were configured as ($200$, $300$), ($300$, $200$), ($167$, $333$), and ($333$, $167$), as shown in Fig.~\ref{fig:mnist_iid_noniid} (d). To accommodate quantum resource constraints, we resized the original $28 \times 28$ images to $4 \times 4$ and encoded each into $4$-qubit states via amplitude embedding. The classification task utilized a $4$-qubit QNN with a three-layer HEA, as shown in Fig.~\ref{fig: framework} (d). Results in Fig.~\ref{fig:mnist_iid_noniid} show rapid convergence of the test loss within the first $40$ communication rounds for both IID and non-IID settings, as depicted in Fig.~\ref{fig:mnist_iid_noniid} (a) and (b). A comparison of final test accuracies across both data distributions, as shown in Fig.~\ref{fig:mnist_iid_noniid} (c), reveals close performance levels (within $1\%$ test accuracy) with a close communication time ($\approx 40$ rounds), highlighting the strong adaptability of our QFL protocol across varying data distribution conditions. \update{Details of the hyperparameter settings are provided in the Supplementary Materials.}

\begin{table}[t]
\centering
\caption{\textbf{Cost of quantum secret keys and test accuracy for $200$ clients under different quantization levels.} } 
\label{table: quantization}
   			\setlength{\tabcolsep}{0.25cm} 
      \renewcommand\arraystretch{1.5}
	\begin{tabular}{c@{\hspace{0.7cm}}c@{\hspace{0.7cm}}c@{\hspace{0.7cm}}}\hline \hline
		&Cost (MB) &  Accuracy \\  \hline
            $32$-bit quantization & $10.593$ & $0.9798$ \\  
		$16$-bit quantization & $5.296$  & $0.9738$ \\ 
		$8$-bit quantization & $2.648$  & $0.9704$ \\ 
            Benchmark & - & $0.9860$ \\  
		\hline\hline
	\end{tabular}
\end{table}

\subsection{Simulation of the large-scale implementation of QuNetQFL}

The QuNetQFL protocol is flexible, supporting both quantum and classical machine learning models. If quantum computational resources are unavailable on either the clients or server, classical models can be used, ensuring scalability while benefiting from quantum-secure communication at scale, particularly in quantum metropolitan area networks.

To evaluate the scalability of QuNetQFL, we implemented it using the classical LeNet5 model ($61706$ parameters) and the MNIST dataset in an IID setting. The training set was equally divided among $200$ clients, each receiving $300$ instances randomly. In each aggregation round, the server randomly selected $5\%$ of the clients. The total number of rounds was set to $200$, and clients performed local training with $5$ local epochs, a batch size of $32$, and the Adam optimizer at a learning rate of $0.01$. Due to limitations in our ability to implement such scale quantum networks, we used pseudo-random numbers as masks instead of true quantum secret keys in this experiment to illustrate the impact of quantization on reducing quantum secret key costs in QuNetQFL.

Table~\ref{table: quantization} presents results for $q = 8, 16, 32$-bit quantization and plaintext aggregation (benchmark), showing the quantum secret key cost (related to model update size) and corresponding test accuracy. The results show that $8$-bit quantization reduces the quantum key cost by $4 \times$, and $16$-bit quantization achieves a $2 \times$ reduction compared to $32$-bit quantization in each communication round. Despite $200$ clients, convergence occurred within $20$ to $40$ rounds, with consistent performance across quantization levels and benchmarks, leading to significant communication savings, lower key costs, and reduced time in real-world collaborative learning. 
This evaluation demonstrates the role of quantization techniques employed in the QuNetQFL for balancing model performance and communication efficiency for scalable quantum-enhanced federated learning.


\section{Discussion}\label{Discussion}
 In this work, we introduce QuNetQFL, a quantum federated learning protocol leveraging quantum networks to achieve information-theoretic security during model updates. By employing QKD, we ensure encrypted communication among clients, where local model updates are protected by quantum secret keys and masked via one-time pads. The protocol also incorporates efficient quantization techniques and is adaptable to other compression methods, such as low-rank tensor compression, weight pruning, and knowledge distillation, to further reduce both communication and quantum key costs.
    
    We experimentally validated QuNetQFL on a five-node quantum network, achieving quantum secret key generation with key rates exceeding $30$ kbps in both $3$-client and $4$-client scenarios.  Notably, protocols such as twin-field QKD, based on the Sagnac loop used in our experimental scheme, can reach distances of up to $200$ km~\cite{mandil2024long}, suggesting the potential for large-scale QFL applications in metropolitan quantum networks. Our protocol demonstrates that near-term quantum networks can already support realistic FL tasks, offering a platform for scaling distributed learning systems in quantum networks. 
    
    Through both simulations and hardware-based evaluations, we demonstrate QuNetQFL's broad applicability across diverse tasks, including quantum resource classification, sentiment analysis, and handwritten digit recognition. These tasks were executed using quantum, classical-quantum hybrid, and classical learning models, each corresponding to different levels of quantum computing requirements. \update{Notably, we fine-tuned a hybrid classical–quantum large language model within our protocol and evaluated it on real quantum hardware for various language-classification tasks. The results showed robust performance, comparable to those obtained in simulation, underscoring the practical viability and scalability of QuNetQFL. This validation extends beyond simulation-based studies, confirming that the protocol can effectively operate in real-world scenarios, even within the context of large language models.}
    
    \update{While this work does not demonstrate a quantum advantage in computational or communication complexity, it provides a practical solution for quantum-secure federated learning. Approaches such as differential privacy degrade model accuracy by introducing noise~\cite{li2021quantum}, while schemes based on homomorphic encryption or post-quantum cryptography with blockchain~\cite{li2024quantum, gharavi2025pqbfl} offer computational security, but rely on unproven hardness assumptions and impose significant computational and communication overheads. Other protocols offering unconditional security, like entanglement-based secure aggregation~\cite{zhang2022federated, wang2024quantum}, require quantum resources beyond current capabilities, and gradient-hiding methods~\cite{li2024privacy} suffer from circuit complexity and accuracy loss. In contrast, QuNetQFL achieves information-theoretic security through QKD-based masking, requiring only pairwise secret keys—resources already available in current quantum networks, making it a practical and near-term route to secure and scalable QFL. \textcolor{black}{We note that the secure-aggregation workflow of QuNetQFL is modular with respect to the key-establishment mechanism. In principle, classical or post-quantum key-establishment methods could also be used to generate the pairwise keys required for cancelling-mask construction, uploading masked local updates, and server-side aggregate recovery. However, such substitutions would change the security properties: QKD-derived keys enable information-theoretic security, whereas classical and post-quantum alternatives provide computational security based on the assumed hardness of the underlying cryptographic primitives.} }

    \update{Additionally, blind quantum computing presents a promising approach to quantum privacy computing, which can seamlessly integrate with our protocol. Blind quantum computation enables clients with minimal quantum resources to delegate tasks to an untrusted server while keeping the input, algorithm, and output confidential~\cite{polacchi2023multi,li2021quantum}. However, existing schemes do not inherently protect the federated learning aggregation phase, where model updates can be vulnerable to gradient-leakage attacks. While differential privacy is used in prior works~\cite{li2021quantum} to protect this phase, it incurs a trade-off in model utility. Our QuNetQFL protocol can replace differential privacy in these schemes, offering information-theoretic security without compromising model accuracy. A detailed algorithm and discussion are provided in the Supplementary Materials. This enhancement positions QuNetQFL to integrate with experimental advancements in multi-user blind quantum computing~\cite{polacchi2023multi}, further advancing the practicality of scalable QFL.}

    Future work will focus on reducing communication complexity by integrating these advanced quantum algorithms, balancing efficiency with practical implementation. \update{Additionally, there is an emerging trend to combine QKD and post-quantum cryptography to harness the strengths of both, enhancing the security and efficiency of quantum communication systems~\cite{wang2021experimental,zeng2024practical}. We believe this approach holds great potential for building privacy-preserving distributed learning systems in the future.}
    Although QuNetQFL was primarily evaluated on variational quantum circuit models for the noisy intermediate-scale quantum era, it is also adaptable to large-scale quantum machine learning models~\cite{liu2024towards}, making it relevant for future fault-tolerant quantum computing.   
    We anticipate that this work will stimulate the development of more practical and scalable QFL schemes, further advancing quantum technologies in distributed learning systems.

\section{Materials and methods}
The detailed protocol for QuNetQFL is provided in Algorithm~\ref{algo:qunetqfl}. Importantly, QuNetQFL is designed to be flexible, allowing clients to choose their local training algorithms based on resource constraints and computational requirements. Clients can employ either gradient-based methods or alternative non-gradient-based methods to update local models, particularly when obtaining the gradients of the QNN circuit is expensive.

\begin{figure}
\begin{algorithm}[H]
\caption{Protocol of QuNetQFL}
\label{algo:qunetqfl}
\begin{algorithmic}
\REQUIRE The untrained global model parameters with  $\boldsymbol{\theta}^0$, The data size distribution of clients $\{n_k \}_{k=1}^K$, the number of iterations $T$, quantization bit length $q$, the specific QKD protocols.  
\ENSURE Trained global model parameters $\boldsymbol{\theta}^*$

\STATE Initialization: Randomly generate a set of $T$ subsets $\{ \mathcal{S}^t\}_{t=1}^T$, where each $\mathcal{S}^t$ contains an equal number of client indices.

\FOR{each round $t \in [T]$}
    \STATE 1) Select the subset of clients' indexes $\mathcal{S}^t$.  
    \STATE 2) For each client $i \in \mathcal{S}^t$:  
        \STATE \hspace{0.5cm} a) Update the local model and compute the local update:
        $\Delta \boldsymbol{\theta}^t_i \gets \boldsymbol{\theta}^t_i - \boldsymbol{\theta}^{t-1}$.
        \STATE \hspace{0.5cm} b) Perform the selected QKD protocol (e.g., \textbf{MDI-QKD}, see Box~1) with each other connected clients in the $\mathcal{S}^t$ in the underlying quantum networks, and generate masking vector: 
        \vspace{-8pt}
        \begin{equation*}
           \mathbf{m}^t_i \gets \sum_{j \in \mathcal{S}^t, j \neq i} (-1)^{i > j}  \text{QK}^t_{i,j} \bmod 2^{q}.
        \end{equation*}
        \vspace{-12pt}
        \STATE \hspace{0.5cm} c) Compute masked and quantized local update:  
         \vspace{-8pt}
        \begin{equation*}
          \Delta \widetilde{\boldsymbol{\theta}}^t_i \gets \left[Q^q(p_i^t \cdot \Delta \boldsymbol{\theta}^t_i) + \mathbf{m}^t_i \right] \bmod 2^q.
        \end{equation*}
        \vspace{-12pt}
    \STATE 3) Aggregate updates:  
        $\Delta\boldsymbol{\theta}^t \gets \left[\sum_{i \in \mathcal{S}^t} \Delta \widetilde{\boldsymbol{\theta}}^t_i \right] \bmod 2^q$.  
    \STATE 4) Update global model:  
        $\boldsymbol{\theta}^t \gets \boldsymbol{\theta}^{t-1} + D^q(\Delta \boldsymbol{\theta}^t)$.  
\ENDFOR

\STATE Output the trained global model parameters $\boldsymbol{\theta}^T$.  
\end{algorithmic}
\end{algorithm}
\end{figure}

    \tcbset{boxrule = 0.5mm, colback=SeaGreen!10!CornflowerBlue!5,colframe=RoyalPurple!55!Aquamarine!100!,
 adjusted title = {\textbf{Box 1\big| MDI-QKD Protocol subroutine}}}
{\begin{tcolorbox}
\label{box:MDI}
    \begin{algorithmic}
        \FOR{Each client $(i, j)$ pairs in $\mathcal{S}^t \times \mathcal{S}^t$ }
            \STATE 1) Randomly choose basis ($X$ or $Y$) and prepare corresponding weak coherent states $\otimes_{k=1}^{n} \ket{e^{\mathbf{i}\boldsymbol{\varphi}_k}\sqrt{\mu}}$, where $\mu$ is the pulse intensity. 
            \STATE 2) Send the states to an untrusted node, Eve, for measurement.
            \STATE 3)  Decide whether to flip the bit according to the measurement outcome.
            \STATE 4) Estimate bit error rate by clients' announcements.
            \STATE 5) Perform postprocessing steps (error correction and privacy amplification) to generate enough final secure keys. 
        \ENDFOR  
    \end{algorithmic}
    \textbf{Note:} 
       Secret keys used in QuNetQFL can be generated during training or pre-generated to the overall process time. MDI-QKD reduces the \update{minimal} quantum resources required for client participation, and QuNetQFL is flexible to emerging QKD techniques. 
\end{tcolorbox}

\subsection{QKD-Based Masking for Secure Aggregation}
In this subsection, we outline the technical details of secure aggregation using QKD-based masking. In each round, client $i \in \mathcal{S}^t$ constructs a mask
\begin{equation}
\mathbf{m}_i^{t}=\sum_{j\in\mathcal{S}^{t},\,j\neq i}(-1)^{\,i>j}\,\mathrm{QK}_{i,j}^{t} \bmod 2^{q},
\end{equation}
where $\mathrm{QK}_{i,j}^{t}$is the pairwise quantum key shared by clients $i$ and $j$. The alternating signs ensure that masks cancel exactly when summed over all participating clients. Client $i$ then uploads the masked, quantized update
\begin{equation}
\label{eq:local_update_mask}
\small
\Delta\widetilde{\boldsymbol{\theta}}_{i}^{t}
=\Big[\,Q^{q}\!\big(p_i^{t}\cdot\Delta\boldsymbol{\theta}_{i}^{t}\big)+\mathbf{m}_i^{t}\,\Big]\bmod 2^{q},\quad
p_i^{t}=\frac{n_i}{\sum_{\ell\in\mathcal{S}^{t}}n_{\ell}},
\end{equation}
where $Q^q(\cdot)$ is the $q$-bit quantization function and $p_i^t$ is the client weight. Here, each $\mathrm{QK}_{i,j}^{t}$ is represented as an $M$-entry vector of $q$-bit words (equivalently, an $(M\times q)$-bit string), with $M$ the model size. Each entry is used as two’s-complement signed integers in $[-(2^{q-1}-1),\,2^{q-1}-1]$, and all masking and aggregation are carried out as additions in $\mathbb{Z}_{2^q}$, rather than the bitwise XOR commonly used with QKD keys in quantum cryptography, to match the $q$-bit quantized updates. 

The server aggregates masked updates by
\begin{align}
\Delta \boldsymbol{\theta}^t 
&= \left[ \sum_{i \in \mathcal{S}^t} \Delta \mathbf{\widetilde{\boldsymbol{\theta}}}^{ t}_{i} \right] \bmod 2^q \\
&= \left[\sum_{i \in \mathcal{S}^t} Q^q(p^t_i \cdot \Delta \boldsymbol{\theta}^t_i) \right] \bmod 2^q,
\end{align}
where the cancellation property
\begin{equation}
\Big[\sum_{i\in\mathcal{S}^t}\mathbf{m}_i^{t}\Big] \bmod 2^q = 0
\end{equation}
ensures that individual masks cancel out and the quantized updates are successfully decrypted.

\subsection{Quantization technique}

To efficiently utilize the secret keys generated in quantum networks and 
further reduce the communication overhead, we employ a quantization technique in QuNetQFL, originally developed for homomorphic encryption~\cite{zhang2020batchcrypt} and later adapted for secure aggregation~\cite{zheng2022aggregation}. This method quantizes a scalar $s \in \mathbb{R}$ within $[-\beta, \beta]$ into a $q$-bit sign integer in the range $[-(2^{q-1}-1), 2^{q-1}-1]$ using the $q$-bit Quantizer $Q^q(s)$:
\begin{equation}
 \label{eq: quan}
    Q^q(s) = \text{sgn}(s) \cdot \text{Round}\left(\text{abs}(s) \cdot (2^{q-1}-1)/ \beta \right),
\end{equation}
where $\text{sgn}(\cdot)$ is the sign function, $\text{abs}(\cdot)$ denotes the absolute value, and $\text{Round}(\cdot)$ maps the input to the nearest integer.
The corresponding de-quantized process for a quantized valued $v$ is given by: 
\begin{equation}
 \label{eq: quan}
    D^q(v) = \text{sgn}(v) \cdot \left(\text{abs}(v) \cdot \beta / (2^{q-1}-1) \right).
\end{equation}
Quantization is performed on the client side for local model updates, while de-quantization is executed on the cloud server for the aggregated global model. Model parameters are clipped to the range $[-\beta, \beta]$ to ensure quantization accuracy. Note that to handle signs during de-quantization, the server employs the following adjustment: If $v > 2^{q-1}-1$, it is updated as $v - 2^q$, otherwise, $v$ remains unchanged.

\textcolor{black}{In practice, $\beta$ is selected according to the dynamic range of the quantity being quantized. For global QNN parameters, which correspond to periodic circuit rotation angles naturally bounded to $[-\pi,\pi]$ by the periodicity of quantum gates, we set $\beta=\pi$. For local model update differences, $\beta$ must additionally prevent integer overflow during modular aggregation in $\mathbb{Z}_{2^q}$ before de-quantization, since the server reconstructs the aggregated quantized update from $N_c$ participating clients rather than individual updates. We therefore conservatively set $\beta = N_c \beta_0$, where $N_c$ is the number of selected clients and $\beta_0$ bounds the typical per-client update magnitude. In our experiments, $\beta_0=1$ for the QNN tasks and $\beta_0=0.1$ for the BERT-QNN fine-tuning tasks, reflecting the smaller update magnitudes in the latter setting. These values were empirically verified in the reported training runs to avoid significant clipping while maintaining stable training accuracy.}

\subsection{Security analysis}

(a) \emph{Threat model.} In our framework, we adopt the standard \textit{honest-but-curious} setting in FL, where clients and the server follow the protocol but may attempt to infer others' data from messages. We exclude the extreme (and impractical) case where the server colludes with all but one client during a secure sum~\cite{bonawitz2017practical}. \update{This \textit{honest-but-curious} assumption also excludes attacks from malicious clients during quantum key exchanges, such as denial-of-service attacks. Using a star-topology QKD network for key distribution could mitigate such risks and move beyond the \textit{honest-but-curious} assumption, representing a valuable direction for future work.} (b) \emph{Security Guarantee.} \update{QKD enables the distribution of unconditional secure keys between two distant parties.} By using pairwise QKD keys as one-time pads in the $q$-bit quantized domain, the uploaded vector $\Delta\widetilde{\boldsymbol{\theta}}_{i}^{t}$ becomes information-theoretically indistinguishable from random to any party lacking the corresponding keys. The server learns only the aggregate (after mask cancellation), never an individual client’s update. Our guarantees protect client data against other clients, the server, and external adversaries. However, protecting the global model is not the goal of this work.

\subsection{Secret-key distribution in experimental quantum network}

To experimentally generate quantum secure keys for demonstrating QuNetQFL with $3$ or $4$ clients, we established a quantum network with five participants. This includes four clients, Alice, Bob, Charlie, and David, who potentially require quantum secret keys for collaborative training, and one party, the untrusted Eve, as shown in Fig.~\ref{fig: exsetup}. \update{In this setup, Eve is responsible for measuring interference results and sending unmodulated coherent pulses as the coherent state source. This can be understood as the "quantum channel" in Fig.~\ref{fig: framework}a at an abstract level.} Through this quantum network, each client pair can securely share quantum secret keys. 

\begin{figure*}[htbp]
    \centering
    \includegraphics[width=1.\textwidth]{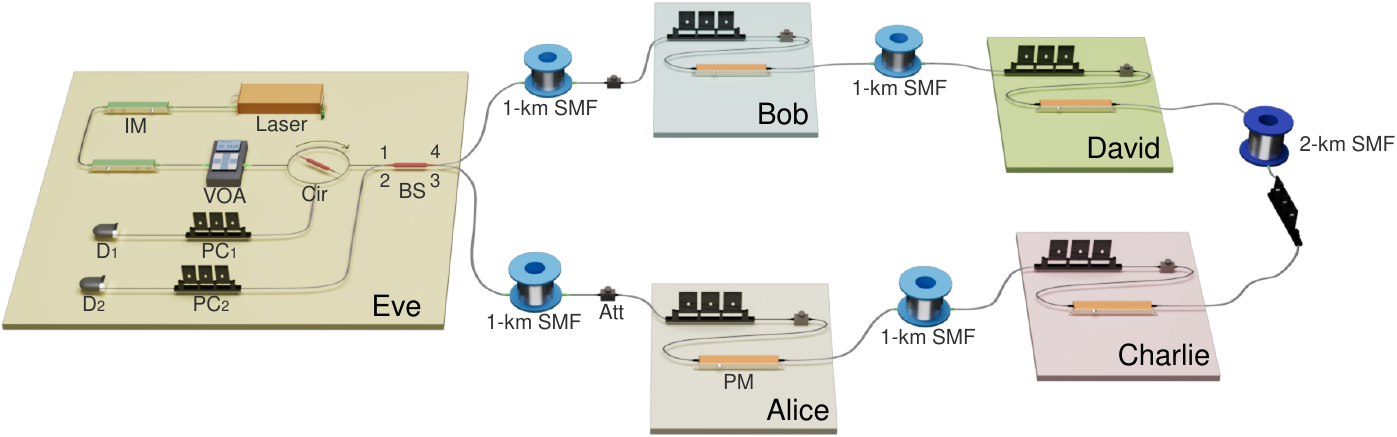}
    \caption{\textbf{Experimental setup of the quantum network.} At Eve’s site, a continuous-wave laser source is employed with two intensity modulators (IM) and a variable optical attenuator (VOA) to produce weak coherent pulses. Four clients (Alice, Bob, Charlie, and David) are interconnected in the Sagnac loop. When any pair of clients need to establish secret keys, Eve injects the pulses into the loop via a circulator (Cir) and a $50:50$ beam splitter (BS). Upon reaching the designated clients, phase modulators (PMs) are employed to add phase to the pulses. After modulation, the two pulse trains interfere at Eve’s BS, where interference results are detected by two superconducting nanowire single-photon detectors ($\rm D_1$ and $\rm D_2$). This structure includes 1-km single-mode fibers (SMFs) between Alice and Eve, Bob and Eve, Alice and Charlie, and Bob and David, while a $2$-km SMF connects Charlie and David. Attenuators (Att) are placed between Alice and Eve, and between Bob and Eve, to adjust channel losses. Polarization controllers (PCs) are used to align polarization in the loop.}
    \label{fig: exsetup}
\end{figure*}

Here, we take the key generation process of MDI-QKD between Alice and Bob as an example, which applies similarly to other client pairs. The system frequency is 100 MHz. The relevant protocol description can be found in Methods.
In Eve’s site, to prepare weak coherent pulses with an extinction ratio exceeding $30$ dB, two intensity modulators are utilized to chop the continuous light from a narrow linewidth continuous-wave laser source (NKT E15). Eve then sends the pulses into the loop after passing through a circulator and a $50:50$ beam splitter (BS). 
In the loop, two clients modulate the pulse train with phase modulators. Note that users only modulate one pulse train. Specifically, Alice only modulates the clockwise pulses and Bob only modulates the counterclockwise pulses. Then, the two pulses interfere at Eve's BS, and Eve records corresponding detection events by $\rm D_1$ and $\rm D_2$.
The detection efficiencies of $\rm D_1$ and $\rm D_2$ are $60.3\%$ and $64.5\%$, respectively, with dark count rates of approximately $18$ Hz and $28$ Hz. The time window length is set to $1.8$ ns. \update{A time window is a fixed-duration interval within each pulse period, during which detection events are considered valid, while detection events outside this window are discarded.} For Alice and Bob, we adjusted the attenuation values by attenuators between the users to make the total channel loss $6$ dB, while for other client pairs, the total channel loss is set to $12$ dB. The attenuation values are adjusted to ensure the channels are symmetric.

Our network design offers enhanced scalability and cost efficiency for QKD. In this setup, a Sagnac loop is employed to stabilize the phases among different users. The laser source and single-photon detectors are positioned at the party, Eve, outside the loop. Regardless of the number of clients, only one Eve is required for detection, thereby reducing the demand for measurement devices \update{when performing MDI-QKD}. Clients in this structure are only required to
perform phase modulation on weak coherent states, eliminating the need for complex quantum state preparation and measurement. This design also allows for easy adjustment of the number of users. Adding or removing a user does not require any modifications to the detection terminal. For users already in the network, when they need to begin the key generation process, they only need to increase or decrease the fiber length to avoid collisions of pulses in both clockwise and counterclockwise directions.

The probability of choosing the $X$ basis is $90\%$ during implementation and the period length of random number is $10000$. Error rates varied across different user pairs, prompting us to use a genetic algorithm to optimize the key rate for each pair under these differing error conditions. This optimization led to variations in pulse intensity across pairs.

\subsection{Protocol of quantum network and its simulation details}

We employed a four-phase MDI-QKD protocol~\cite{gu2022experimental} between each pair of participants to generate quantum secret keys. To illustrate this process, we use the example of Alice and Bob with an untrusted third party, Eve. All other user pairs in the network follow this same protocol to obtain their quantum secret keys.

(1) Alice and Bob choose the $X$ basis with the probability $p_x$ and the $Y$ basis with the probability $p_y$ ($p_y = 1 - p_x$). For the $X$ basis, they prepare $\ket{e^{\mathbf{i}k_x\pi}\sqrt{\mu}}$ where $k_x$ is the logic bit value ($k_x \in \{0, 1\}$) and $\mu$ is the pulse intensity. For the $Y$ basis, they prepare  $\ket{e^{\mathbf{i}(k_y+1/2)\pi}\sqrt{\mu}}$ where $k_y$ is the logic bit value ($k_y \in \{0, 1\}$).

(2) Eve performs measurements on the pulses with a $50:50$ BS and two single-photon detectors and records the detection results. One and only one detector clicks is defined as an effective event. 

(3) The above steps are repeated many times to accumulate sufficient data. Eve announces all effective events and the corresponding detector that clicks. For each effective event announced by Eve, if $\rm D_2$ clicks, Bob flips his corresponding logic bit. Alice and Charlie retain only the logical bits from effective measurements, discarding others.
Then they disclose their basis choices for effective events through authenticated classical channels.

(4) Alice and Bob announce all their bit values in the $Y$ basis to calculate the quantum bit error
rate $E_b^y$ in the $Y$ basis to estimate the phase error rate under the $X$ basis $E_{\rm p}$, and the number of counts $n_x$ and $n_y$ can also be obtained
under $X$ and $Y$ bases, respectively.

(5) Alice and Bob perform error correction on the remaining keys under the $X$ basis and privacy amplification to obtain the final secret keys.

The final key rate of MDI-QKD~\cite{gu2022experimental} can be given by

\begin{equation}\label{finite}
	\begin{aligned}
		l = \ &n_{x}[1 -H(\overline{E}_{\rm p})]-{\lambda}_{\rm EC}\\
        &-{\rm log}_2\frac{2}{\epsilon_{\rm EC}} 
        -{\rm log}_2\frac{1}{4\epsilon^2_{\rm PA}},
	\end{aligned}
\end{equation}
where $n_x$ is the number of total counts in the $X$ basis and $\overline{E}_{\rm p}$ is the upper bound of phase error rate under the $X$ basis. $\lambda_{\rm EC} = n_xfH(E_b^x)$ is the leaked information during error correction, where $f$ is the error correction efficiency, $E_{\rm b}^x$ is the bit error rate under the $X$ basis and $H(x) = -x\log_2x - (1-x)\log_2(1-x)$ denotes the binary Shannon entropy. $\epsilon_{\rm EC}$ and $\epsilon_{\rm PA}$ are the failure probabilities for the error correction and privacy amplification, respectively and we set $\epsilon_{\rm EC} = \epsilon_{\rm PA} = 10^{-10}$. 

$E_{\rm b}^x$ can be obtained from experimental results and $E_{\rm p}$ is bounded by the following inequality
\begin{equation}\label{epbound}
	1-2\Delta\leq\sqrt{E_{\rm b}^{y}E_{\rm p}}+\sqrt{(1-E_{\rm b}^{y})(1-E_{\rm p})},
\end{equation}
where $\Delta = \left(1-|\braket{\Psi_{Y}|\Psi_{X}}|^2\right)/2Q$. $Q =n_{\rm tot}/N $ is the total gain, where $n_{\rm tot}$ is the number of detection events and $N$ is the number of pulses sent. $\ket{\Psi_{X}}$ $\left(\ket{\Psi_{Y}}\right)$ is the basis-dependent state under the $X$ ($Y$) basis. The fidelity can be expressed as
\begin{equation}\label{practicalfinal}
	\begin{aligned}
		\braket{\Psi_{X}|\Psi_{Y}}=&\frac{1}{4}\lbrack(1-\mathbf{i})\braket{\sqrt \mu|\mathbf{i}\sqrt \mu}+(1-\mathbf{i})\\
        &\braket{-\sqrt \mu|-\mathbf{i}\sqrt \mu}+(1+\mathbf{i})\braket{\sqrt \mu|-\mathbf{i}\sqrt \mu}\\
        &+(1+\mathbf{i})\braket{-\sqrt \mu|\mathbf{i}\sqrt \mu}\rbrack.
	\end{aligned}
\end{equation}
Considering the finite-key effect, Kato's inequality~\cite{kato2020concentration} is utilized. The upper bound of the expectation value $m_y^{*}$ is given by $m_y + \Delta_{n_y}$, where $m_y = n_y E_{\rm b}^y$ represents the number of errors in the $Y$ basis and $\Delta_{n_y} = \sqrt{\frac{1}{2}n_y{\rm ln}\epsilon_{\rm F}^{-1}}$ with the failure probability $\epsilon_{\rm F} = 10^{-10}$. 
${E_{\rm b}^{y}}^{*}=m_y^{*}/n_y$ can be calculated and $E_{\rm p}^{*}$ can be derived according to Eq.~\ref{epbound}. Thus we can get the number of phase errors $m_{\rm p}^{*}=n_x E_{\rm p}^{*}$ and $\overline{m}_{\rm p}$ can subsequently be estimated by the inequality. Consequently, we can get the upper bound of phase error rate $\overline{E}_{\rm p}=\overline{m}_{\rm p}/n_x$. \update{The security of four-phase MDI-QKD is presented in Ref.~\cite{gu2022experimental}.}

\subsection{Two quantum datasets}

In this subsection, we introduced the details of two quantum datasets used to evaluate the performance of QuNetQFL in entanglement classification and nonstabilizerness classification tasks. Entanglement is a fundamental resource for quantum information processing tasks, especially in quantum communication. A given $n$-qubit state $\ket{\psi}$ is a product state if and only if $\ket{\psi} = \otimes_{i = 1}^n \ket{\psi_i}$, otherwise, it is entangled. For this task, we employed the NTangled dataset~\cite{schatzki2021entangled} which quantifies multipartite entanglement using Concentratable Entanglement (CE), defined as $C(\ket{\psi}) = 1 - \frac{1}{2^n}\sum_{\alpha \in P} \text{Tr}[\rho_{\alpha}^2]$. Here, $P$ is the power set of the set $\{1,2,...,n\}$, and the $\rho_{\alpha}$ is the reduced density operator with respect to the index $\alpha$. States with higher $C(\ket{\psi})$ exhibit greater entanglement. Following the methods in Ref.~\cite{schatzki2021entangled}, we generated a balanced $3$-qubit dataset consisting of low ($\text{CE} = 0.05$) and high ($\text{CE} = 0.35$) entangled states using a six-layer HEA. In an IID setting, each client accessed $160$ states for training, while the server processed $200$ test states with equal proportions from the two classes in both $3$-client and $4$-client scenarios.

Quantum magic, or nonstabilizerness~\cite{Howard2017,WWSM19}, is crucial for achieving quantum computational advantage beyond classical simulation~\cite{gottesman1997stabilizer}. Stabilizer states, generated by Clifford operations, are classically simulable, whereas non-stabilizer states possess quantum magic. For this task, we employed a popular magic measure, \textit{stabilizer R\'enyi entropy}~\cite{Leone_2022}, to quantity nonstabilizeness. A balanced $3$-qubit dataset was generated, comprising non-stabilizer states with \textit{stabilizer R\'enyi entropy} greater than $1.5$ (sampled Haar-randomly) and stabilizer states selected from the total of $1080$ three-qubit stabilizer states. Each client accessed $120$ training states, and the server processed $120$ test states, with equal class proportions in both $3$-client and $4$-client scenarios. 

For both classification tasks, \update{we assume all clients are quantum-enabled, equipped with quantum computing devices that can implement QNNs, while the server needs only to aggregate updates in classical parameter form and can therefore be entirely classical. However, the protocol is also compatible with a quantum server if needed to validate the performance of the aggregated quantum model at each step, potentially enabling early termination of the protocol when appropriate.} We used QNNs with the HEA, as shown in Fig.~\ref{fig: framework} (c). To enhance the capacity of the learning model, we input the states parallelly twice and adopted a $6$-qubit HEA with $4$ layers. The label predictions were obtained by measuring the last qubit of the circuit on the $Z$ basis. \textcolor{black}{For the binary classification tasks considered here, a single readout qubit is sufficient to encode the prediction score after the variational circuit, while also reducing the measurement overhead on near-term quantum hardware.} These two tasks were performed with $16$-bit quantization ($q = 16$) over $200$ and $160$ communication rounds, respectively. Each client conducted local training with a batch size of $32$, an initial learning rate of $0.01$ and $0.02$ (for the two tasks, respectively), and the Adam optimizer, employing mean squared error as the loss function, with one local epoch per round. 

\subsection{Federated fine-tuning a hybrid BERT–QNN model}

In this subsection, we introduce the methods of federated fine-tuning a hybrid classical-quantum model, leveraging the power of pre-trained classical models for transfer learning. We use the large language model Bert as a pre-trained classical model to extract complex features, which are then processed on a quantum device through a QNN. When federated traning such a model to solve real-world language processing tasks, we freeze the parameters of BERT and fine-tuning only the parameters of the QNN and the connecting linear layer. This approach allows us to leverage the power of a pre-trained classical model while enhancing its performance with quantum processing. In the sentiment analysis task, the BERT-QNN architecture consists of BERT, followed by a linear layer that reduces the $768$-dimensional BERT output to a $4$-dimensional vector, and a $4$-qubit, three-layer QNN using a HEA. In this setup, only $3112$ parameters are fine-tuned, compared to the total over one hundred million parameters of the BERT model. We note that our experimentally generated secret keys are sufficient to support this model size, with the number of communication rounds set to $100$.
\textcolor{black}{For comparison, we also implemented a dimension-matched classical baseline in which the QNN module was replaced by a lightweight classical nonlinear classification head while keeping the frozen BERT encoder, the $4$-dimensional bottleneck, the four-client IID setting, $100$ communication rounds, $16$-bit quantization, and QKD-key-based secure aggregation unchanged. This baseline contains $3081$ trainable parameters.}
Details of the simulated setup are provided in the Supplementary Materials.

\subsection*{Acknowledgments}
This work was supported by the National Natural Science Foundation of China (No. U25D8016, No. 12522419, No. 12274223), the Fundamental Research Funds for the Central Universities and the Research Funds of Renmin University of China (No. 24XNKJ14), the Program for Innovative Talents and Entrepreneurs in Jiangsu (No. JSSCRC2021484), and the China Postdoctoral Science Foundation (No. 2025M783395).


%

\makeatother



\onecolumngrid

\begin{center}
\large{\textbf{Supplementary Information:}\\ \textbf{
Experimentally validated quantum-secure federated learning \\over a multi-user quantum network}}
\end{center}

\section{Convergence analysis }
\label{appendix: theo_analysis}
In this section, we provide a theoretical convergence analysis for our quantum federated learning (QFL) framework, named QuNetQFL, under the use of a gradient-based optimization algorithm. The analysis begins with the introduction of four key assumptions and concludes with Theorem~\ref{theo: convergence_ana}, which formalizes the convergence behavior of the framework.

Consider round $t$ of our federated learning process, where the global model parameters are denoted by $	\boldsymbol{\theta}^t \in \mathbb{R}^M$. At each round, $K$ clients are selected to update the model locally. Each client receives the global model and performs $	\tau$ iterations of local gradient descent, starting from $	\boldsymbol{\theta}_k^{t,0} = 	\boldsymbol{\theta}^t$. The local update for $k$-th client is given by:

\begin{align}
\boldsymbol{\theta}_k^{t,s + 1} = \boldsymbol{\theta}_k^{t,s} - \eta g(\boldsymbol{\theta}_k^{t,s}; \xi_k^{t,s}), \quad s = 0, 1, \ldots, \tau - 1,
\end{align}
where $\eta$ is the learning rate and $\xi_k^{t,s}$ denotes the mini-batch of data sampled from the $k$-th client's local data at iteration $s$. After $\tau$ iterations, client $k$ obtains the updated parameters $\boldsymbol{\theta}_k^{t, \tau}$. Next, the client computes the parameter difference $\Delta \boldsymbol{\theta}_k^t = \boldsymbol{\theta}_{k}^{t,\tau} - \boldsymbol{\theta}^t$, clips it to the range $[- \beta, \beta ]$, and quantizes it into $q$ bits. Next, the client then adds a privacy mask $\mathbf{m}_k^t$ and sends the masked update $\Delta \mathbf{\widetilde{\boldsymbol{\theta}}}^{ t}_{k} = \left[ Q^q(p_k \Delta \boldsymbol{\theta}^t_k) + \mathbf{m}_k^t \right] \bmod 2^q$ to the server.

On the server side, the aggregated model update is computed as follows:

\begin{align}
   \boldsymbol{\theta}^{t+1} 
    &= \boldsymbol{\theta}^t +  D^q \left( \sum_{k=1}^K \Delta \mathbf{\widetilde{\boldsymbol{\theta}}}^{ t}_{k}\right) \\ \notag
    &= \boldsymbol{\theta}^t + D^q \left( \sum_{k=1}^K [Q^q(p_k\Delta\boldsymbol{\theta}_k^t) + \mathbf{m}_k^t ]\bmod 2^q\right)
\end{align}
where $ p_k = \frac{n_k}{N} $, with $ n_k $ denoting the number of local data samples held by the $k$-th client, and $ N $ representing the total number of data samples across all participating clients. Here, $ Q^q(\cdot) $ is the $q$-bit quantization function, and $ D^q(\cdot)$ is the corresponding $q$-bit dequantization function. The masking terms $\mathbf{m}_k^t$ does not affect the final aggregated result, since $[\sum_k \mathbf{m}_k^t] \bmod 2^q = 0$.

The objective is to solve the following optimization problem:

\begin{equation}
\min_{\boldsymbol{\theta} \in \mathbb{R}^M} f(\boldsymbol{\theta}) = \sum_{k=1}^K p_k f_k(\boldsymbol{\theta}), 
\end{equation}
where $f_k(\boldsymbol{\theta}) = \mathbb{E}_{\xi \sim \mathcal{D}_k}[f_k(\boldsymbol{\theta}; \xi)]$ is the loss function of the $k$-th client, and $\mathcal{D}_k$ represents the local data distribution for client $k$. 

Next, we outline the assumptions typically used in the convergence analysis of federated learning~\cite{10185969,amiri2020federated, NEURIPS2018_17326d10,bubeck2015convex}. 

    \begin{assumption}
    \label{assump1}
    
    The loss function $f_k$ is $\mu$-strongly convex and satisfies the Polyak-Lojasiewicz (PL) inequality \cite{karimi2016linear}:  
    \[
    2\mu(f_k(\boldsymbol{\theta}) - f^*) \leq \|\nabla f_k(\boldsymbol{\theta})\|_2^2,
    \]  
    for any $\boldsymbol{\theta}$, where $f^*$ denotes the global optimum.
    
    \end{assumption}
    
    \begin{assumption} 
    \label{assump2}
    
    The loss function $f_k$ is $L$-smooth with a Lipschitz constant $L > 0$. That is, for any $\boldsymbol{\theta}_1, \boldsymbol{\theta}_2$, the following inequality holds:  
    \[
    f_k(\boldsymbol{\theta}_2) \leq f_k(\boldsymbol{\theta}_1) + \nabla f_k(\boldsymbol{\theta}_1)^{T}(\boldsymbol{\theta}_2 - \boldsymbol{\theta}_1) + \frac{L}{2} \| \boldsymbol{\theta}_2 - \boldsymbol{\theta}_1 \|^2.
    \]

    \end{assumption}

    \begin{assumption}
    \label{assump3}
    
        The stochastic gradient $g(\boldsymbol{\theta}_{k}^{t,s}) = \nabla f(\boldsymbol{\theta}_{k}^{t,s})$ is unbiased and bounded. Specifically,  
        \[
        \mathbb{E}_{\xi \sim \mathcal{D}_k}[g(\boldsymbol{\theta}_{k}^{t,s})] = \nabla f_k(\boldsymbol{\theta}_{k}^{t,s}), \quad \text{and} \quad \text{Var}(\nabla f(\boldsymbol{\theta}_{k}^{t,s})) \leq \sigma^2,
        \]  
         where the mini-batch gradient $g(\boldsymbol{\theta}_{k}^{t,s})$ is defined as $g(\boldsymbol{\theta}_k^{t,s}) = \frac{1}{b} \sum_{j=1}^b \nabla f(\boldsymbol{\theta}_k^{t,s}; \xi_{k}^{t,s,j})$ and $b$ is the mini-batch size.    
        Based on these assumptions, the expected squared norm of the mini-batch gradient is bounded as follows:  
        \[
        \mathbb{E}[\| g(\boldsymbol{\theta}_k^{t,s}) \|^2] \leq \| \nabla f_k(\boldsymbol{\theta}_k^{t,s}) \|^2 + \frac{\sigma^2}{b}.
        \]  
    
    \end{assumption}

    \begin{assumption}
    \label{assump4}
    For a parameter $\Delta\boldsymbol{\theta}_m$ in $\Delta\boldsymbol{\theta}_k^t$, the quantized result $Q(\Delta\boldsymbol{\theta}_m)$ satisfies: $Q(\Delta\boldsymbol{\theta}_m) = \Delta\boldsymbol{\theta}_m + \omega$, where $\omega$ is a random variable distributed as $\omega \sim U(-\delta_q, \delta_q)$ with $\delta_q = \frac{\beta}{2^{q-1} - 1}$, and $U (\cdot)$ denotes the uniform random distribution. 
    Thus, for the entire parameter $\Delta\boldsymbol{\theta}_k^t$, the quantized result is \(Q(\Delta\boldsymbol{\theta}_k^t) = \Delta\boldsymbol{\theta}_k^t + \Omega\), where \(\Omega\) is an \(M\)-dimensional vector, with each component independently distributed as \(U(-\delta_q, \delta_q)\).
    \end{assumption}

    \begin{theorem}
    \label{theo: convergence_ana}
    Under the assumptions above, if the server and clients update the global model $\boldsymbol{\theta}^t \in \mathbb{R}^M$ following the described protocol, and the learning rate $\eta$ satisfies $\eta \leq \frac{1}{L}$, the expected convergence of the global loss function after $T$ rounds of training is bounded by:
    \begin{align}
    \mathbb{E}[f(\boldsymbol{\theta}^T)] - f^* 
    \leq
    (1-\eta\mu)^{T\tau}(\mathbb{E}[f(\boldsymbol{\theta}^0)] - f^*) 
    + [1-(1-\eta\mu)^{T\tau}][E_g + \frac{E_q}{1-(1-\eta\mu)^\tau}]
    \end{align}
    where $E_g = \frac{\kappa\eta \sigma^{2}}{2b}$ with $\kappa = L/\mu$ being the condition number, and $E_q = \frac{LMK^2\delta_q^2}{6}$.
    \end{theorem}

    \begin{proof}
    
    The global model aggregated by the sever at the $(t+1)$-th round is given as:
    \begin{align}
       \boldsymbol{\theta}^{t+1} 
         &= \boldsymbol{\theta}^t + D^q \left( \sum_{k=1}^K \left[Q^q(p_k\Delta\boldsymbol{\theta}_i^t) + \mathbf{m}_k^t \right]\bmod 2^q\right) \notag\\ 
         &= \boldsymbol{\theta}^t + \sum_{k=1}^K (p_k \Delta\boldsymbol{\theta}_k^t + \Omega) \notag\\ 
         &=\sum_{k=1}^K p_k\boldsymbol{\theta}_k^{t,\tau} + K\cdot\Omega
    \end{align}
where the sum of the masking item equals zero, and the second equation comes from the Assumption~\ref{assump4}.

    Define $\overline{\boldsymbol{\theta}}^{t+1} = \sum_{k=1}^K p_k \boldsymbol{\theta}_k^{t,\tau}$ ,we have
    \begin{align}
        \mathbb{E}[\boldsymbol{\theta}^{t+1}] = \sum_{k=1}^{K} p_k \boldsymbol{\theta}_k^{t,\tau} =\overline{\boldsymbol{\theta}}^{t+1}
    \end{align}

    Using Assumption~\ref{assump2} and set $\boldsymbol{\theta}_2= \boldsymbol{\theta}^{t+1} $ and $ \boldsymbol{\theta}_1 = \overline{\boldsymbol{\theta}}^{t+1}$, we have
    \begin{equation}
        f(\boldsymbol{\theta}^{t+1}) \leq f(\overline{\boldsymbol{\theta}}^{t+1}) + \nabla f(\overline{\boldsymbol{\theta}}^{t+1})^{T}(K\cdot\Omega)+\frac{L}{2}\Vert K\cdot\Omega \Vert ^{2}
    \end{equation}
    Since $\mathbb{E}[\boldsymbol{\theta}^{t+1}] = \overline{\boldsymbol{\theta}}^{t+1}$, and $K\cdot\Omega$ is independent of $\nabla F(\overline{\boldsymbol{\theta}}^{t+1})$, taking the expectation gives:

\begin{align}
    \label{s4}
    \mathbb{E}[f(\boldsymbol{\theta}^{t+1})]
    &\leq \mathbb{E}[f(\overline{\boldsymbol{\theta}}^{t+1})] + \frac{L}{2} \mathbb{E}[\Vert K\cdot\Omega \Vert ^{2}]\notag \\ 
    &= \mathbb{E}[f(\overline{\boldsymbol{\theta}}^{t+1})]  + \frac{L}{2}\frac{MK^2\delta_q^2}{3}   
\end{align}

As the loss function $f$ is convex, we have 
\begin{align}    
    f(\overline{\boldsymbol{\theta}}^{t+1}) = f(\sum_{k=1}^K p_k \boldsymbol{\theta}_k^{t,\tau}) \leq \sum_{k=1}^{K} p_k f_k(\boldsymbol{\theta}_k^{t,\tau}) 
\end{align}
and then
\begin{align}
    \label{s6}
    \mathbb{E}[f(\overline{\boldsymbol{\theta}}^{t+1})]  \leq \sum_{k=1}^{K} p_k \mathbb{E}[f_k(\boldsymbol{\theta}_k^{t,\tau}) ]
\end{align}
Locally, the client updates the parameters with stochastic gradient descent. Specifically, $\boldsymbol{\theta}_k^{t,s+1} = \boldsymbol{\theta}_k^{t,s} - \eta g(\boldsymbol{\theta}_{k}^{t,s};\xi_{k}^{t,s})$. 
Using Assumption~\ref{assump2}, for client k 
\begin{align}
    f_k(\boldsymbol{\theta}_{k}^{t,s+1}) - f_k(\boldsymbol{\theta}_{k}^{t,s})
    &\leq \nabla f_k(\boldsymbol{\theta}_{k}^{t,s})^{\top}(\boldsymbol{\theta}_k^{t,s+1} -         \boldsymbol{\theta}_k^{t,s}) + \frac{L}{2}||\boldsymbol{\theta}_k^{t,s+1} - \boldsymbol{\theta}_k^{t,s}||^{2} \notag\\ 
    &\leq -\eta \nabla f_k(\boldsymbol{\theta}_k^{t,s})^{\top}g(\boldsymbol{\theta}_k^{t,s}; \xi_k^{t,s}) + \frac{L}{2}||-\eta g(\boldsymbol{\theta}_k^{t,s}; \xi_k^{t,s})||^{2} \notag\\
    &=-\eta \nabla f_k(\boldsymbol{\theta}_k^{t,s})^{\top}g(\boldsymbol{\theta}_k^{t,s}; \xi_k^{t,s}) + \frac{L\eta^{2}}{2} ||g(\boldsymbol{\theta}_k^{t,s}; \xi_k^{t,s})||^{2} 
\end{align}

After taking expectations on both sides and using Assumption~\ref{assump3}, what we have is the following:
\begin{align}
\mathbb{E}[f_k(\boldsymbol{\theta}_k^{t,s+1}) - f_k(\boldsymbol{\theta}_k^{t,s})]  \leq -\eta ||\nabla f_k(\boldsymbol{\theta}_k^{t,s})||^{2} + \frac{L\eta^{2}}{2}||\nabla f_k(\boldsymbol{\theta}_k^{t,s})||^{2} + \frac{L\eta^{2} \sigma^{2}}{2b}  
\end{align}

After simplification, we obtain:
\begin{align}
    \mathbb{E}[f_k(\boldsymbol{\theta}_k^{t,s+1}) - f_k(\boldsymbol{\theta}_k^{t,s})] = (\eta-\frac{L\eta^{2}}{2}) (-||\nabla f_k(\boldsymbol{\theta}_k^{t,s})||^{2}) + \frac{L\eta^{2} \sigma^{2}}{2b}    
\end{align}

If we choose $\eta \leq \frac{1}{L}$, then
\begin{align}
\mathbb{E}[f_k(\boldsymbol{\theta}_k^{t,s+1})] - \mathbb{E}[ f_k(\boldsymbol{\theta}_k^{t,s})] \leq \frac{\eta}{2}(-||\nabla f_k(\boldsymbol{\theta}_k^{t,s})||^{2}) + \frac{L\eta^{2} \sigma^{2}}{2b}  
\end{align}

According to  \ref{assump1}, we have: 
\begin{align}
    \Vert\nabla f_k(\boldsymbol{\theta}_k^{t,s})\Vert^2 \geq 2 \mu (f_k(\boldsymbol{\theta}_k^{t,s} ) - f^*)
\end{align}

As a result, 
\begin{align}
    \mathbb{E}[f_k(\boldsymbol{\theta}_k^{t,s+1})] - \mathbb{E}[ f_k(\boldsymbol{\theta}_k^{t,s})] \leq  -\eta\mu (\mathbb{E}[f_k(\boldsymbol{\theta}_k^{t,s} ) ]- f^*) +  \frac{L\eta^{2} \sigma^{2}}{2b}
\end{align}

By transforming the formula and iterating it for $\tau$ steps, while using $\boldsymbol{\theta}_k^{t,0} = \boldsymbol{\theta}^{t}$, we obtain:
\begin{align}
    \label{s9}
    \mathbb{E}[f_k(\boldsymbol{\theta}_k^{t,\tau})] - f^{*} - \frac{L\eta \sigma^{2}}{2\mu b} \leq (1-\eta\mu)^{\tau}(\mathbb{E}[f_k(\boldsymbol{\theta}^{t})]-f^{*} - \frac{L\eta \sigma^{2}}{2\mu b}) 
\end{align}

Sum up all clients, we have
\begin{align}
    \label{s11}
    \sum_{k=1}^{K} p_k \mathbb{E}[f_k(\boldsymbol{\theta}_k^{t,\tau})] - f^{*} - \frac{L\eta \sigma^{2}}{2\mu b} \leq (1-\eta\mu)^{\tau}(\mathbb{E}[f(\boldsymbol{\theta}^{t}))]-f^{*} - \frac{L\eta \sigma^{2}}{2\mu b}) 
\end{align}

Combine Eq.~\eqref{s4}, Eq.~\eqref{s6} and Eq.~\eqref{s11}, it is obvious that
\begin{align}
    \mathbb{E}[f(\boldsymbol{\theta}^{t+1})]&\leq \sum_{k=1}^{K} p_k \mathbb{E}[f_k(\boldsymbol{\theta}_k^{t,\tau})]  +\frac{LMK^2\delta_q^2}{6}                 \notag \\
    &=(1-\eta\mu)^{\tau}(\mathbb{E}[f(\boldsymbol{\theta}^{t})]-f^{*} - \frac{L\eta \sigma^{2}}{2\mu b})  +f^{*} + \frac{L\eta \sigma^{2}}{2\mu b}+\frac{LMK^2\delta_q^2}{6}
\end{align}

We denote the error item caused by batch gradient descent as $E_g = \frac{\kappa\eta \sigma^{2}}{2 b}$ with $\kappa = L/\mu$ being the condition number, and $\frac{LMK^2\delta_q^2}{6}$ as the error $E_q$ caused by quantization. Therefore, we have the following relationship:
\begin{align}
    \mathbb{E}[f(\boldsymbol{\theta}^{t+1})]- f^* - \left[E_g + \frac{E_q}{1-(1-\eta\mu)^\tau}\right]
    \leq 
    (1-\eta\mu)^{\tau}\left(\mathbb{E}[f(\boldsymbol{\theta}^{t})]-f^{*} - \left[E_g + \frac{E_q}{1-(1-\eta\mu)^\tau}\right]\right)
\end{align}
After performing $T$ rounds of training globally, the following equation can be obtained:
\begin{align}
    \mathbb{E}[f(\boldsymbol{\theta}^{T})]- f^* - \left[E_g + \frac{E_q}{1-(1-\eta\mu)^\tau}\right]
    \leq 
    (1-\eta\mu)^{T\tau}\left(\mathbb{E}[f(\boldsymbol{\theta}^{0})]-f^{*} - \left[E_g + \frac{E_q}{1-(1-\eta\mu)^\tau}\right]\right)
\end{align}

Then we arrive at the conclusion that
\begin{align}
    \label{final}
    \mathbb{E}[f(\boldsymbol{\theta}^T)] - f^* 
    \leq
    (1-\eta\mu)^{T\tau} \left(\mathbb{E}[f(\boldsymbol{\theta}^0)] - f^* \right) 
    + \left[1-(1-\eta\mu)^{T\tau}\right] \left[E_g + \frac{E_q}{1-(1-\eta\mu)^\tau} \right]
\end{align}

\end{proof}

\textbf{Asymptotic Convergence Analysis}: In summary, the model converges to the optimal solution \( f^* \) over time, with the first term in the equation representing the exponential decay of the initial error. However, due to stochastic noise and quantization errors, the second term introduces a constant error floor. As a result, the error is upper-bounded by $E_g + \frac{E_q}{1-(1-\eta\mu)^\tau}$ as \( T \to \infty \). Specifically, by adopting strategies to reduce the condition number $\kappa$ and selecting an appropriate quantization precision relative to the model size, this asymptotic upper bound can be made small enough. Moreover, this asymptotic convergence analysis indicates restricting the number of selected clients in each round can accelerate the convergence of training in our QFL framework. This analysis not only provides a theoretical convergence guarantee but also offers practical guidance for training in the QuNetQFL framework, further highlighting its practical advantages.


\section{Quantum Neural Network Model}

In this section, we briefly introduce the Quantum Neural Network (QNN) model~\cite
{cerezo2021variational}, which uses a series of parameterized quantum gates to simulate classical neural network learning. Given a training dataset $\{\mathbf{x}_i, y_i\}_{i=1}^N$, classical input data $\mathbf{x}_i \in \mathbb{R}^d$ is encoded into an $n$-qubit quantum state $\rho(\mathbf{x}_i) = U_{\text{emb}}(\mathbf{x}_i) \ket{0}^{\otimes n} \bra{0}^{\otimes n} U_{\text{emb}}^\dagger(\mathbf{x}_i)$ in a QNN, where $U_{\text{emb}}(\mathbf{x}_i)$ is a unitary embedding operator. The common encoding strategies include amplitude encoding, and angle encoding~\cite{PhysRevA.103.032430, larose2020robust}. The quantum state is processed by a series of parameterized quantum gates $U(\boldsymbol{\theta})$, with $\boldsymbol{\theta}$ as the trainable parameters. 

The output is measured using an observable operator $O$, and the model’s prediction $E(\boldsymbol{\theta})$ is the expectation value of the observable:
\begin{equation}
    E(\boldsymbol{\theta}) = \text{Tr} \left[ U(\boldsymbol{\theta}) U_{\text{emb}}(\mathbf{x}_i) \ket{0}^{\otimes n} \bra{0}^{\otimes n} U_{\text{emb}}^\dagger(\mathbf{x}_i) U^\dagger(\boldsymbol{\theta}) O \right]
\end{equation}
QNNs are trained by minimizing a loss function $f(\boldsymbol{\theta})$, which quantifies the difference between the predicted outputs $E(\boldsymbol{\theta})$ and actual labels $y$. The loss function is optimized using gradient-based techniques, with gradients evaluated via the parameter shift rule~\cite{mitarai2018quantum}.



\newpage

\section{Sentiment analysis using Hybrid BERT–QNN}
In this section, we outline the specific parameter settings of numerical experiments for sentiment analysis using Hybrid BERT–QNN in Table~\ref{tabel_bert_qnn}. 

\update{\begin{table*}[ht]
\centering
\caption{\textbf{Parameter settings for Sentiment analysis using Hybrid BERT–QNN}}
\label{tabel_bert_qnn}
\setlength{\tabcolsep}{0.3cm} 
\renewcommand\arraystretch{1.5}
\begin{tabular}{c@{\hspace{1.5cm}}c@{\hspace{1.5cm}}c@{\hspace{1.5cm}}c@{\hspace{1.5cm}}}
\hline \hline 
\textbf{Parameter}          & \textbf{MC} (\textbf{RP}) & \textbf{Yelp}(\textbf{IMDb})  & \textbf{Amazon} \\ \hline
Training set (per client)   & 17  & 300 & 300    \\         
Test set (Server)           & 30    & 300  & 300   \\    
Optimizer                   & Adam  & Adam & Adam\\ 
Learning rate               & 0.05  & 0.05 & 0.05 \\  
Batch size                  & 64 & 128  & 32\\ 
Local epoch                 &  100  &  16 & 16\\ 
Loss function               & MSE   & MSE  & MSE\\ 
\hline \hline
\end{tabular}
\end{table*}}

\newpage

\section{Evaluation on Quantum neural networks classifying MNIST}
In this section, we outline the specific parameter settings of numerical experiments for classifying MNIST via QNNs in Table~\ref{tabel_mnist}. 

\begin{table*}[ht]
\centering
\caption{\textbf{Parameter settings for classifying MNIST via QNN}}
\label{tabel_mnist}
\setlength{\tabcolsep}{0.3cm} 
\renewcommand\arraystretch{1.5}

\begin{tabular}{c@{\hspace{1.5cm}}c@{\hspace{1.5cm}}c@{\hspace{1.5cm}}}
\hline \hline
\textbf{Parameter}          & \textbf{Non-IID} (\textbf{IID})\\ \hline

Training set (per client)   & 500   \\               
Test set (Server)           & 500   \\               
Optimizer                   & Adam  \\
Learning rate               & 0.01  \\    
Batch size                  & 50 \\
Local epoch                 &  1  \\
Loss function               & MSE   \\ 
\hline \hline
\end{tabular}
\end{table*}

\newpage
\section{Evaluation of the large-scale implementation of QuNetQFL}
In this section, we present Fig.~\ref{fig:MNIST_Q}, which illustrates the test accuracy and loss over $200$ communication rounds for $200$ clients, evaluated at different quantization levels. Despite the involvement of up to $200$ clients in the collaborative learning process on the MNIST dataset, we observe a rapid convergence within approximately $20$ to $40$ rounds, with consistent performance across various quantization levels and the benchmark. This demonstrates the efficiency and scalability of QuNetQFL in real-world datasets, achieving rapid convergence and consistent performance even in large-scale, distributed quantum-enhanced learning environments.

\begin{figure*}[ht]
    \centering
    \includegraphics[width=0.85\linewidth]{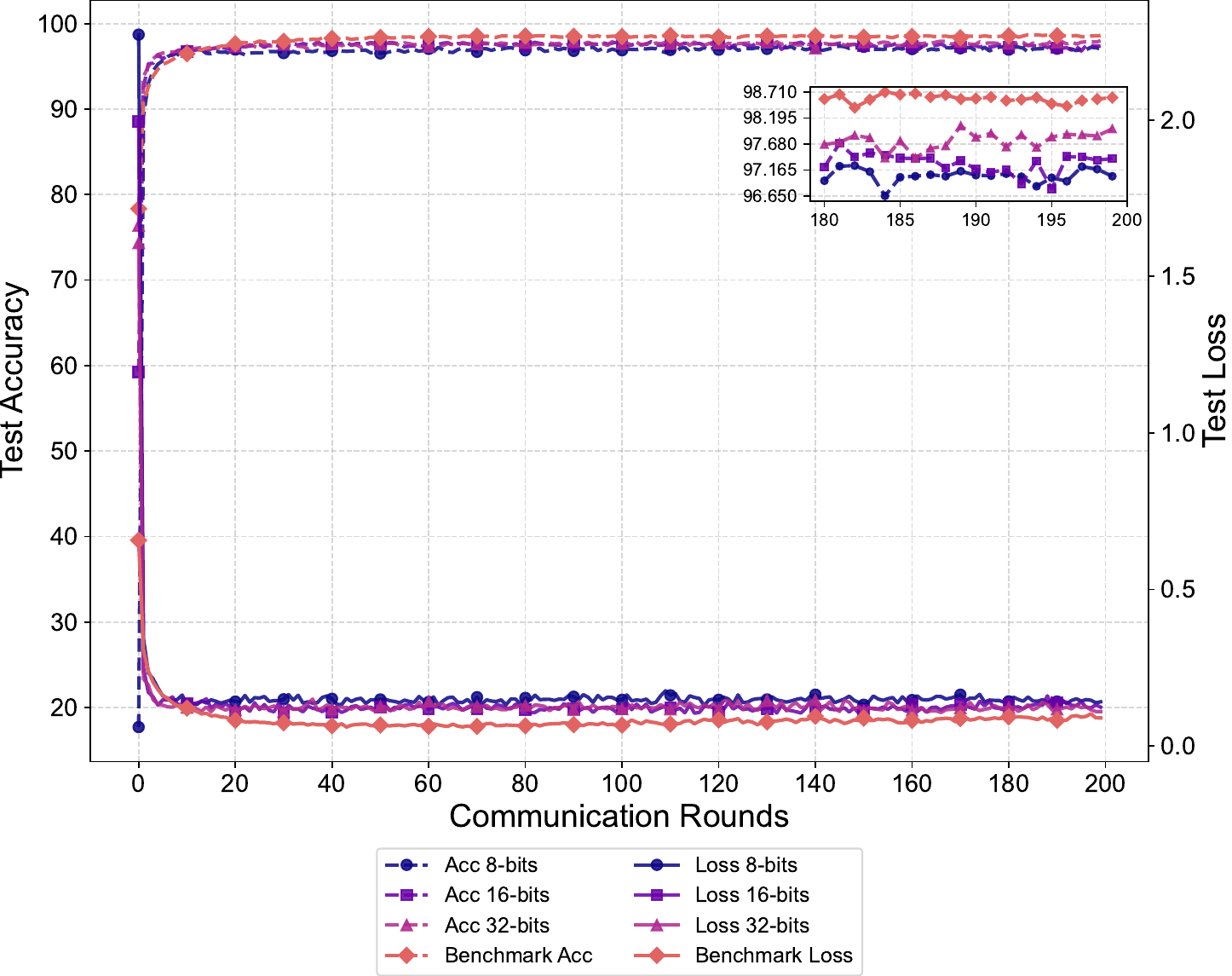}
    \caption{Test accuracy and loss across $200$ communication rounds for $200$ clients, evaluated at different quantization levels.}
    \label{fig:MNIST_Q}
\end{figure*}


\section{Error correction and privacy amplification.}
In this section, we present the details of error correction and privacy amplification techniques used in our experimental quantum network. To obtain the final keys from the raw keys, error correction is required to correct bit errors, and privacy amplification is needed to eliminate the possibility of information leakage.

The raw keys of the participants are transformed into fully correlated bit strings by error correction, utilizing the Cascade key agreement mechanism~\cite{martinez2013key}. Initially, the participants divide their keys into blocks with a variable length, determined by the estimated bit error rate. Based on the recommendation of~\cite{ye2010information} and our optimization, we designated the block length as $0.7/E_{\rm b}^x$ with a growth factor of $2$ and performed a total of $4$ rounds.

In the first round, the participants calculate and compare the parity check values for each block.  Subsequently, they use a dichotomy method to rectify blocks with errors. Following these procedures, each block will only contain an even number (including $0$) of error bits. 

In the following rounds, the participants first shuffle their keys with pre-shared random numbers, then proceed with block division and dichotomy error correction. After identifying and correcting new error bits, the participants backtrack their bit strings by applying the inverse sequence of random numbers and rectify blocks from previous rounds that corresponding to error bits. This is necessary because, after correcting an erroneous bit, the corresponding blocks from previous rounds will contain an odd number of errors, requiring additional error correction. Therefore, the additional error correction in the corresponding blocks is achievable. Through the above methods, we have achieved a stable error correction coefficient of $f<1.2$ for our experimental data according to~\cite{ye2010information}.

After error correction, the keys of the participants become identical. However, the risk of information leakage remains. A privacy amplification process is required to compress the information and ensure confidentiality.

Our privacy amplification algorithm is based on the fast Fourier transform (FFT)  and inverse fast Fourier transform (IFFT)~\cite{liu2016fit}. The final key length $r$ is determined through finite key analysis and the length of raw keys $n$. We use an random binary $(n-1)$-bit string $[v_0, v_1, \cdots, v_{n-2}]$ to construct a $r\times (n-r)$ Toeplitz matrix $V_{r\times (n-r)}$ and combine an $r\times r$ identity matrix with it horizontally to obtain a modified matrix 
\begin{equation}
S_{r\times n} = \begin{bmatrix}
1 &   &   &   &  & v_{r-1} & v_r & \cdots & v_{n-2}\\
  & 1 &   &   &  & v_{r-2}  & v_{r-1} &   & v_{n-3}\\
  &   & \ddots  &   &  &\vdots  &   & \ddots  & \vdots\\
  &   &   & 1 &  & v_1  &   &   & v_{n-r}\\
  &   &   &   & 1 & v_0  & v_1 & \cdots & v_{n-r-1}\\

\end{bmatrix}. 
\end{equation}
Afterwards, $X_n = [x_0, x_1, \cdots, x_{n-1}]'$ is defined as the vector of the raw keys and $Y_r = [y_0, y_1, \cdots, y_{r-1}]'$ is defined as the vector of the final keys. There is:

\begin{equation}
Y_r = S_{r\times n}\times X_n = X_r + Y_r',
\end{equation}
where  $Y_r'$ can be obtain by $D_{n}\otimes X_{n}'$~\cite{li2019high}. 
Eq.~\ref{dn} represents a circulant matrix $D_{n}$ extended from the Toeplitz matrix $V_{r\times (n-r)}$. The vector $X_{n}' = [0,\cdots, 0 , x_r, \cdots, x_{n-1}]$ is obtained by setting the first $r$ values of the vector $X_{n}$ to zero. The calculation procedure is presented in Eq.~\ref{fft}.

\begin{equation}\label{dn}
D_{n} = \begin{bmatrix}
P_{(n-r)\times r} & P_{(n-r)\times (n-r)} \\
P_{r \times r} & V_{r\times (n-r)}
\end{bmatrix},
\end{equation} 

\begin{equation}\label{fft}
\begin{aligned}
&\begin{bmatrix}
 P_{(n-r)\times (n-r)} \times X_{n-r}\\
Y_r'
\end{bmatrix} \\
& = D_{n}\otimes X_{n}' \\
& =\mathrm{IFFT[FFT}(D_{n}) \cdot \mathrm{FFT}(X_{n}')].
\end{aligned}
\end{equation}

The privacy amplification method described above is applied to the experimental data, integrating the final key vector $Y_r$, resulting in a processing rate exceeding $10$ Mbps.

\newpage
\section{Detailed experimental data of quantum networks}
\label{appendix: detailed experimental data}
In this section, we present the calibrated efficiencies of the devices in Eve's site in Table~\ref{table_loss}. The insertion losses of four users (Alice, Bob, Charlie, and David) are $3.62$ dB, $3.08$ dB, $3.16$ dB, and $3.01$ dB, respectively. Table~\ref{table_3user} records the corresponding data used for key rate calculation.

\begin{table}[ht]
\centering
\caption{\textbf{The efficiencies of devices.} } 
\label{table_loss}
   			\setlength{\tabcolsep}{0.2cm} 
      \renewcommand\arraystretch{1.5}
	
	\begin{tabular}{c@{\hspace{1.5cm}}c@{\hspace{1.5cm}}}\hline \hline
		Optical element &  Efficiency\\  \hline
		Cir 2$\rightarrow$3  & 89.3\% \\ 
		BS-1  & 84.7\% \\ 
		BS-2  & 85.5\% \\ 
		$\rm{PC_1}$   & 92.7\% \\
		$\rm{PC_2}$   & 90.1\% \\ \hline\hline
	\end{tabular}
\end{table}

\begin{table}[ht]
\centering
\caption{\textbf{Summary of experimental data.} 
The total number of detector clicks, $n_{tot}$, includes the clicks under the 
$X$ basis ($n_x$), the $Y$ basis ($n_y$), and the detection events corresponding to phase differences that are odd multiples of $\pi/2$.
$m_x$ ($m_y$) represents the number of detection errors under the $X$ ($Y$) basis.
$E_{\rm b}^x$ and $E_{\rm b}^y$ represent the bit error rate under the $X$ basis and under the $Y$ basis, respectively. $\lambda_{\rm EC}$ is the leaked information during error correction. A, B, C, and D represent Alice, Bob, Charlie, and David, respectively.} 
\label{table_3user}
   			\setlength{\tabcolsep}{0.1cm} 
      \renewcommand\arraystretch{1.5}
	
	\begin{tabular}{ccc ccc ccc cc}\hline \hline
            & Client pair&Intensity&$n_{tot}$&$n_x$&$m_x$&$n_y$ & $m_y$ & $E_{\rm b}^x$ & $E_{\rm b}^y$ & $\lambda_{\rm EC}$ \\ \hline
		\multirow{3}{*}{3-client}&AB & 0.017 &  208796444 &  169216602 &1434989& 2171543& 10748 & 0.85\%&0.49\% & 14220217  \\ 
            &AC & 0.0085 & 52455918 & 42542127 &446184 &575833&7221   &1.05\% &1.25\% &4219418  \\ 
            &AD & 0.0089  &  54706564 & 44275485&437014 &517003&3405   &0.99\% & 0.66\% &4221496 \\
            \hline
            
            \multirow{6}{*}{4-client}&AB & 0.017 & 209641454&169711875 &1301843&2095785 &8263
            & 0.77\% & 0.39\% & 13122399 \\ 
            
            &AC & 0.0083 & 51270791 &41489668 &463434 &472642 &6228    &1.11\% & 1.32\% &4378680 \\ 
            &AD & 0.0087  & 53621226&43467119 &439500 &536452 &4636     &1.01\% & 0.86\% &4234674\\
            &BC & 0.0087 & 53175349 &43089366 &429297 &561109 &4106     &1.00\% & 0.73\% &4139771\\ 
            &BD & 0.0087 & 53520583 & 43268936 &478145 &456832 &5182     & 1.10\% & 1.13\% &4520879\\ 
            &CD & 0.0074  & 45406632 &36791065 &530094 &442536 &5523  &1.44\% &1.25\% & 4720904 \\

  \hline\hline
	\end{tabular}
\end{table}

\newpage
\update{\section{Integrating QuNetQFL with Blind Quantum Computation}}

\update{
In this section, we integrate QuNetQFL with Blind Quantum Computation (BQC) to establish a practical, secure framework for distributed quantum learning. QuNetQFL leverages quantum key distribution (QKD) to implement quantum One-time pad masking (OTPM) for the secure exchange of model updates during the aggregation phase of federated learning. Meanwhile, BQC~\cite{broadbent2009universal, fitzsimons2017private} enables clients with minimal quantum capabilities to delegate computations to untrusted servers without revealing their private data. By incorporating BQC into QuNetQFL, we not only enhance the privacy of model aggregation but also significantly reduce the quantum resource requirements on each participant, such as local quantum devices to perform QNNs.}

\update{Existing QFL protocols through BQC~\cite{li2021quantum}, such as those based on universal blind quantum computation~\cite{broadbent2009universal}, often rely on differential privacy to hide updates by adding noise. However, this approach can degrade model performance and does not ensure perfect secrecy. In contrast, we merges QuNetQFL protocol with BQC to achieve \textbf{information-theoretic security} without compromising performance. In our framework, each client computes its local model update and secures it using quantum secret keys distributed via QKD network. The detailed algorithm for integrating QuNetQFL with BQC is presented in Algorithm~\ref{algo:qunetqfl_bqc}, which slightly modifies the original QuNetQFL protocol described in the manuscript.}


\begin{figure}
\begin{algorithm}[H]
\caption{\update{QuNetQFL Protocol with BQC}}
\label{algo:qunetqfl_bqc}
\begin{algorithmic}
\REQUIRE The untrained global model parameters with  $\boldsymbol{\theta}^0$, The data size distribution of clients $\{n_k \}_{k=1}^K$, the number of iterations $T$, quantization bit length $q$, the specific QKD protocols. 
\ENSURE Trained global model parameters $\boldsymbol{\theta}^*$

\STATE Initialization: Randomly generate a set of $T$ subsets $\{ \mathcal{S}^t\}_{t=1}^T$, where each $\mathcal{S}^t$ contains an equal number of client indices.

\FOR{each round $t \in [T]$}
    \STATE 1) Select the subset of clients' indexes $\mathcal{S}^t$.  
    \STATE 2) For each client $i \in \mathcal{S}^t$:  
    
        \STATE \hspace{0.5cm} a) Update the local model and compute the local update through the \textbf{UBQC protocol}:
        $\Delta \boldsymbol{\theta}^t_i \gets \boldsymbol{\theta}^t_i - \boldsymbol{\theta}^{t-1}$.
        \STATE \hspace{0.5cm} b) Perform a \textbf{specifical QKD protocol} with each other connected clients in the $\mathcal{S}^t$ in the underlying full-connected quantum networks, and generate masking vector: 
        \vspace{-8pt}
        \begin{equation}
           \mathbf{m}^t_i \gets \sum_{j \in \mathcal{S}^t, j \neq i} (-1)^{i > j} \cdot \text{QK}^t_{i,j}.
        \end{equation}
        \vspace{-12pt}
        \STATE \hspace{0.5cm} c) Compute masked and quantized local update:  
         \vspace{-8pt}
        \begin{equation}
          \Delta \widetilde{\boldsymbol{\theta}}^t_i \gets \left[Q^q(p_i^t \cdot \Delta \boldsymbol{\theta}^t_i) + \mathbf{m}^t_i \right] \bmod 2^q.
        \end{equation}
        \vspace{-12pt}
    \STATE 3) Aggregate updates:  
        $\Delta\boldsymbol{\theta}^t \gets \left[\sum_{i \in \mathcal{S}^t} \Delta \widetilde{\boldsymbol{\theta}}^t_i \right] \bmod 2^q$.  
    \STATE 4) Update global model:  
        $\boldsymbol{\theta}^t \gets \boldsymbol{\theta}^{t-1} + D^q(\Delta \boldsymbol{\theta}^t)$.  
\ENDFOR

\STATE Output the trained global model parameters $\boldsymbol{\theta}^T$.
\end{algorithmic}
\end{algorithm}
\end{figure} 





\end{document}